\documentclass[aip,amsmath,amssymb,reprint]{revtex4-1}
\usepackage{graphicx}
\usepackage{dcolumn}
\usepackage{bm}
\usepackage[utf8]{inputenc}
\usepackage[T1]{fontenc}
\usepackage{mathptmx}
\usepackage{etoolbox}
\makeatletter
\def\@email#1#2{%
 \endgroup
 \patchcmd{\titleblock@produce}
  {\frontmatter@RRAPformat}
  {\frontmatter@RRAPformat{\produce@RRAP{*#1\href{mailto:#2}{#2}}}\frontmatter@RRAPformat}
  {}{}
}%
\makeatother
\begin{document}
\preprint{AIP/123-QED}
\title{From the Rose\textendash DuBois Ansatz of Hot Spot Fields to the Instanton Solution: a Pedestrian Presentation}
\author{Philippe Mounaix}
\email{philippe.mounaix@polytechnique.edu}
\affiliation{CPHT, CNRS, \'Ecole
polytechnique, Institut Polytechnique de Paris, 91120 Palaiseau, France.}
\date{\today}
\begin{abstract}
This paper gives a pedestrian presentation of some technical results recently published in mathematical physics with non-trivial implications for laser-plasma interaction. The aim is to get across the main results without going into the details of the calculations, nor offering a specialist's user guide, but by focusing conceptually on how these results modify the commonly-held description -- in terms of laser hot spot fields -- of backscattering instabilities with a spatially smoothed laser beam. The intended readers are plasma physicists as well as graduate students interested in laser-plasma interaction. No prior knowledge of scattering instabilities is required. Step by step, we explain how the laser hot spots are gradually replaced with other structures, called {\it instantons}, as the amplification of the scattered light increases. In the amplification range of interest for laser-plasma interaction, instanton\textendash hot spot complexes tend to appear in the laser field (in addition to the expected hot spots), with a non-negligible probability. For even larger amplifications and systems longer than a hot spot length, the hot spot field description is clearly invalidated by the instanton takeover.
\end{abstract}
\maketitle
%
%
\section{Introduction}\label{intro}
The results reported here are the fruits of a reassessment of the assumption underlying the Rose and DuBois's 1994 seminal paper\cite{RD1994} entitled "Laser Hot Spots and the Breakdown of Linear Instability Theory with Application to Stimulated Brillouin Scattering". A good way of introducing our presentation is to explain the different elements of the title of Ref.~\onlinecite{RD1994}. This is the subject of this section at an elementary level. The interested reader will find technical accessible introductions to laser-plasma interaction in, e.g., Refs.~\onlinecite{SG1969,Drake1974,Kruer1988,Michel2023}.
\subsection{Stimulated Brillouin scattering}\label{sbs}
\begin{figure}[!b]
\includegraphics[width=0.9\linewidth]{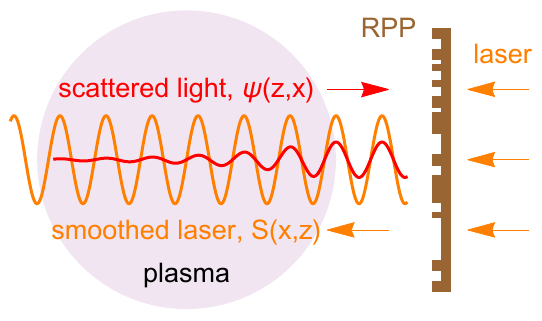}
\caption{\label{figure1}Schematic representation of a backscattering instability of a spatially smoothed laser beam in the strongly damped regime considered by Rose and DuBois in Ref.~\onlinecite{RD1994}.}
\end{figure}
The context is laser-plasma interaction in the configuration schematically represented in Fig.~\ref{figure1}. The laser coming from the right, passes through a random phase plate (RPP) that breaks its spatial coherence\cite{Kato1984}. As a result, the laser field downstream of the RPP, denoted by $S(x,z)$, is {\it smoothed} (i.e. random on small scales, much smaller than the beam diameter). As the laser propagates through the plasma, part of its light, denoted by $\psi(x,z)$, is scattered back in the opposite direction and amplified. Clearly, this type of scattering is not the usual kind in which light disperses in all directions without amplification: it is called {\it stimulated scattering}. The term Brillouin indicates that the process is accompanied by the emission of a sound wave in the same direction as the laser\cite{SG1969,Drake1974,Kruer1988,Michel2023}. As this sound wave plays no role in what follows, we will ignore it and, for our purposes, "stimulated Brillouin scattering" in the title of Ref.~\onlinecite{RD1994} refers to the process depicted in Fig.~\ref{figure1}, where $S$ and $\psi$ are complex amplitudes.
\subsection{Linear instability theory}\label{lit}
There can be different regimes of scattering, depending on laser and plasma parameters. The regime considered in Ref.~\onlinecite{RD1994} is the one in which $\psi(x,z)$ is the solution to
\begin{equation}\label{withDeq}
\left\lbrace
\begin{array}{l}
\partial_z\psi(x,z)-\frac{i}{2m}\nabla_{x}^2 \psi(x,z)=g\vert S(x,z)\vert^2\psi(x,z), \\
0\le z\le L,\ x\in\Gamma\subset\mathbb{R}^d,\ {\rm and\ given}\ \psi(x,0)\not\equiv 0,
\end{array}\right.
\end{equation}
where $z$ and $x$ respectively denote the coordinates along and perpendicular to the scattered light direction in a plasma of length $L$ and cross-sectional domain $\Gamma$, with $d=1$ or $2$. The real parameters $m\ne 0$ and $g>0$ are proportional to the scattered light wave number and average laser intensity, respectively. In Eq.~(\ref{withDeq}), the laser field $S(x,z)$ is entirely determined by the RPP, with damping and nonlinear effects during propagation through the plasma assumed negligible. This equation is the "linear instability theory" in the title of Ref.~\onlinecite{RD1994}. For the reader familiar with laser-plasma interaction, it corresponds to the convective, or strongly damped, linear regime of stimulated Brillouin backscattering. In this regime, the sound wave is enslaved to the laser and scattered light, which explains why it disappears from the problem. Note that the same Eq.~(\ref{withDeq}) (with different $m$ and $g$) holds for the strongly damped linear regime of stimulated Raman scattering, where the sound wave is replaced with a plasma wave\cite{SG1969,Drake1974,Kruer1988,Michel2023}.

For $g=0$, Eq.~(\ref{withDeq}) reduces to a paraxial Helmholtz equation describing the diffraction of $\psi(x,z)$ along $x$. Setting $g>0$ switches on its amplification along $z$, turning Eq.~(\ref{withDeq}) into a linear stochastic amplifier. Linear because the equation is linear in $\psi(x,z)$, and stochastic because $S(x,z)$ is a random field. In the limit of many RPP elements (the small square teeth on the RPP in Fig~\ref{figure1}), $S(x,z)$ tends to a Gaussian random field by application of the central limit theorem. For technical convenience, this limit is routinely assumed in the analytical part of the theoretical studies on the subject, and $S(x,z)$ is a complex, homogeneous, Gaussian field with $\langle S\rangle=\langle S^2\rangle=0$ and $\langle SS^\ast\rangle$ depending on the RPP energy spectrum\cite{RD1993}. Finally, the standard definition of $2g$ as the power gain exponent at average laser intensity imposes the normalization $\langle\vert S(x,z)\vert^2\rangle=1$.
\subsection{The breakdown of linear instability theory}\label{boflit}
\begin{figure}[!b]
\includegraphics[width=0.8\linewidth]{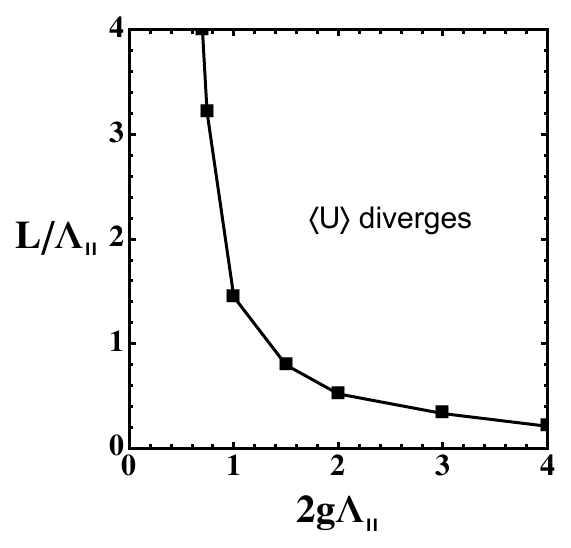}
\caption{\label{figure2}Regions in the $(g,\, L)$ plane where $\langle U\rangle$ is well defined (below the curve), and where $\langle U\rangle$ diverges (above the curve). Adapted from http://dx.doi.org/10.1103/PhysRevLett.72.2883 with permission.}
\end{figure}
Figure~\ref{figure2} reproduces one of the important results of Ref.~\onlinecite{RD1994}. Let $U$ denote the amplification defined by $U=\vert\psi(0,L)\vert^2$ where, for a given realization of $S(x,z)$, $\psi(x,z)$ is determined by solving Eq.~(\ref{withDeq}) numerically, with $\psi(x,0)=1$. Since $S(x,z)$ is a random field, $U$ is a random variable. Averaging $U$ over a large number of realizations, one obtains the results summarized in Fig.~\ref{figure2}, where $\langle\cdot\rangle$ denotes the average over the realizations of $S(x,z)$ and $\Lambda_\parallel$ is the correlation length of $S(x,z)$ along $z$. The curve separates the parameter space $(g,\, L)$ into a region where $\langle U\rangle$ is well defined (below the curve), and a region where numerical results indicate that $\langle U\rangle$ diverges (above the curve). Note that in the limit referred to in Ref.~\onlinecite{RD1994} as the independent hot spot model, this divergence was pointed out by Akhmanov {\it et al.} 20 years before\cite{ADP1974}.

Physically (and numerically), there is of course no divergence. What the divergence of $\langle U\rangle$ means is the occurence of an {\it intermittent} behavior of the amplification in the sense that, for a large (finite) sample of $U$s, the sample average $\overline{U}$ is dominated by the few largest values of $U$ in the sample. If the sample is big enough, these values can be so large that the linear theory of Eq.~(\ref{withDeq}) does not apply to the dominating realizations, even though it remains perfectly valid for all the other realizations in the sample. This is what "the breakdown of linear instability theory" in the title of Ref.~\onlinecite{RD1994} means: it is the breakdown of the linear theory for the few realizations that dominates $\overline{U}$ when $\langle U\rangle$ diverges.

The parameter space region where $\langle U\rangle$ diverges defines the supercritical regime $g>g_c(L)$, where the critical coupling $g_c(L)$ is the smallest $g$ for which $\langle U\rangle =\infty$ at fixed $L$. A finite $g_c(L)$ indicates the existence of two distinct scattering regimes: one characterized by an intermittent nonlinear amplification, in the supercritical regime ($g>g_c(L)$), and the other by the mean linear amplification, $\langle U\rangle <\infty$, in the complementary subcritical regime ($g<g_c(L)$). It is thus important to know the value of $g_c(L)$. Since numerical results are necessarily finite they cannot see the divergence of $\langle U\rangle$, and any numerical estimate of $g_c(L)$ must be inferred from an inevitably not fully controllable extrapolation. It follows in particular that the critical curve in Fig.~\ref{figure2} is an approximation whose deviation from the exact one is not easy to estimate. To get the exact value of $g_c(L)$ the problem must be dealt with analytically. This is a highly non-trivial problem which has been solved in Ref.~\onlinecite{MCL2006}, the result of which will be used in Sec.~\ref{test} for checking the validity of the $g_c(L)$ obtained from the instanton analysis in Sec.~\ref{inst}.

Write $p(U)$ the probability distribution of $U$. Whether or not $\langle U\rangle$ diverges depends on the upper tail of $p(U)$, which in turn depends on the value of $g$. In the supercritical regime, $p(U)$ is a slowly decreasing (fat-tailed) distribution and $\langle U\rangle$ diverges. In this case and according to the discussion above, the realizations of $S(x,z)$ that determine $\overline{U}$ are the ones for which $U$ is large, in the upper tail of $p(U)$. Since it is in these realizations that nonlinear effects associated with the supercritical breakdown of the linear theory are expected to grow, it is essential to be able to identify them. This brings us to the last element of the title of Ref.~\onlinecite{RD1994} to explain.
\subsection{Laser hot spots and the Rose\textendash DuBois ansatz}\label{lhs}
For typical values of the amplification in the bulk of $p(U)$, the realizations of $S(x,z)$ are speckle patterns with local maxima of $\vert S(x,z)\vert^2$ uniformly distributed over the whole interaction region. Maxima greater than about $3\langle\vert S\vert^2\rangle$ are called {\it hot spots} and we will refer to such speckle patterns as {\it hot spot fields}. In this range of $U$, realizations of $S(x,z)$ different from hot spot fields are so rare that they are completely negligible, from which it follows that typical realizations of $S(x,z)$ are hot spot fields. It can be shown\cite{RD1993} that the characteristic hot spot length and width are respectively of the order of $\Lambda_\parallel$ and $\Lambda_\perp$, the correlation length of $S(x,z)$ along $x$. Hot spots are therefore localized structures of the laser field.

On the other hand, if $U$ is in the upper tail of $p(U)$, say above the 95th percentile, $U_{95\%}$, the corresponding realizations of $S(x,z)$ are rare, or atypical, and as such have no a priori reason to resemble typical realizations. At this point, Rose and DuBois make a fairly natural assumption: they assume that realizations in the tail of $p(U)$ {\it are} hot spot fields similar to typical realizations except for the presence of a few hot spots with anomalously high intensity. According to this idea, (i) what makes a realization atypical is the very high intensity of a few hot spots in an otherwise normal hot spot field; (ii) it is the amplification of $\psi(x,z)$ in these rare and intense hot spots that puts $U$ in the tail of $p(U)$ and causes the divergence of $\langle U\rangle$ for $g>g_c(L)$; and therefore (iii) high-intensity hot spots dominate the supercritical regime.

The assumption made by Rose and DuBois serves as an ansatz to develop a practically applicable theory of laser-plasma interaction in the supercritical regime. Throughout the rest of the paper, the terms {\it hot spot field description} and {\it Rose\textendash DuBois ansatz} will be used interchangeably.
\begin{figure}[!b]
\includegraphics[width=0.9\linewidth]{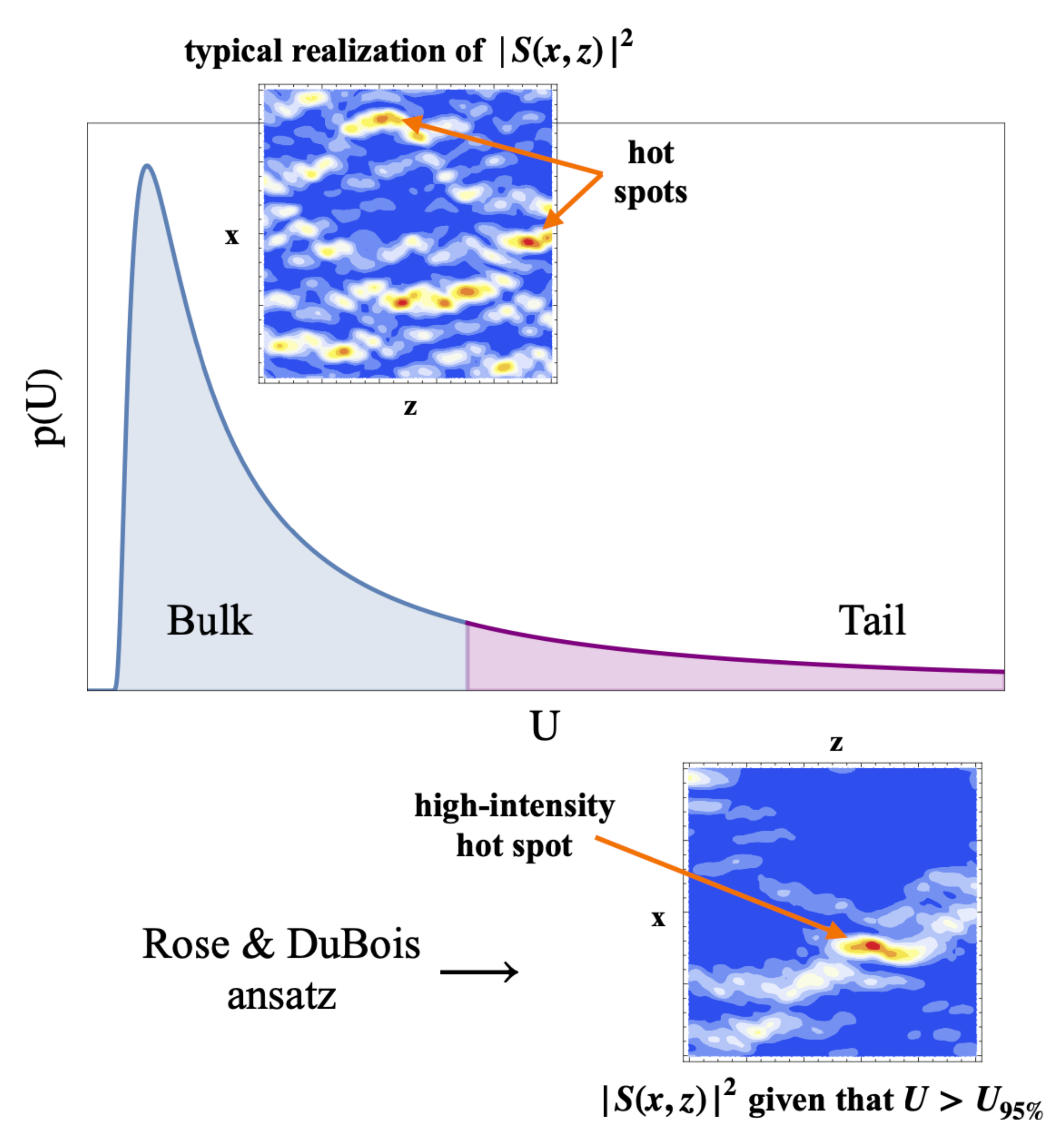}
\caption{\label{figure3}Illustration of the Rose\textendash DuBois ansatz: (left) typical realizations of $\vert S(x,z)\vert^2$ in the bulk of $p(U)$ are hot spot fields; (right) realizations in the tail of $p(U)$ are assumed to be hot spot fields with a few high-intensity hot spots.}
\end{figure}

Figure~\ref{figure3} provides an at-a-glance understanding of the Rose\textendash DuBois ansatz. The hot spot fields displayed in the figure are realizations of $\vert S(x,z)\vert^2$ for $S(x,z)$ given in Eq.~(\ref{solSnum}). The seemingly flat appearance of $\vert S(x,z)\vert^2$ outside the high-intensity hot spot in the realization in the tail of $p(U)$ (compared to the one in the bulk) is due to the color scale being set to the largest value. This makes the normal fluctuations of $\vert S(x,z)\vert^2$ appear smaller than they are. According to the Rose\textendash DuBois ansatz, there are no structures other than hot spots in the laser field. For the realization associated with the tail of $p(U)$ in Fig.~\ref{figure3}, the large value of $U$ is entirely attributed to the amplification of $\psi$ in the high-intensity hot spot.
\subsection{The problem to be solved}\label{tptbs}
Surprisingly enough, to the author's knowledge, the assumption that realizations in the tail of $p(U)$ are hot spot fields has never been tested or even discussed. It is worth quoting the only passage in Ref~\onlinecite{RD1994} where the question is mentioned: "[the laser field $S(x,z)$] inherits the spatial variation of intensity which is either a fully three-dimensional RPP global hot spot field or an isolated hot spot." There is no doubt about that for typical realizations in the bulk of $p(U)$, but for atypical realizations in the tail of $p(U)$, the assertion should be supported by at least some evidence. Until this is done, it is a conjecture, not an established fact true for all $U$. The somewhat misleading affirmative formulation in Ref~\onlinecite{RD1994} should not blind us to the fact that, for the time being, the Rose\textendash DuBois ansatz is an a priori assumption waiting to be justified.

\begin{figure}[!b]
\includegraphics[width=0.9\linewidth]{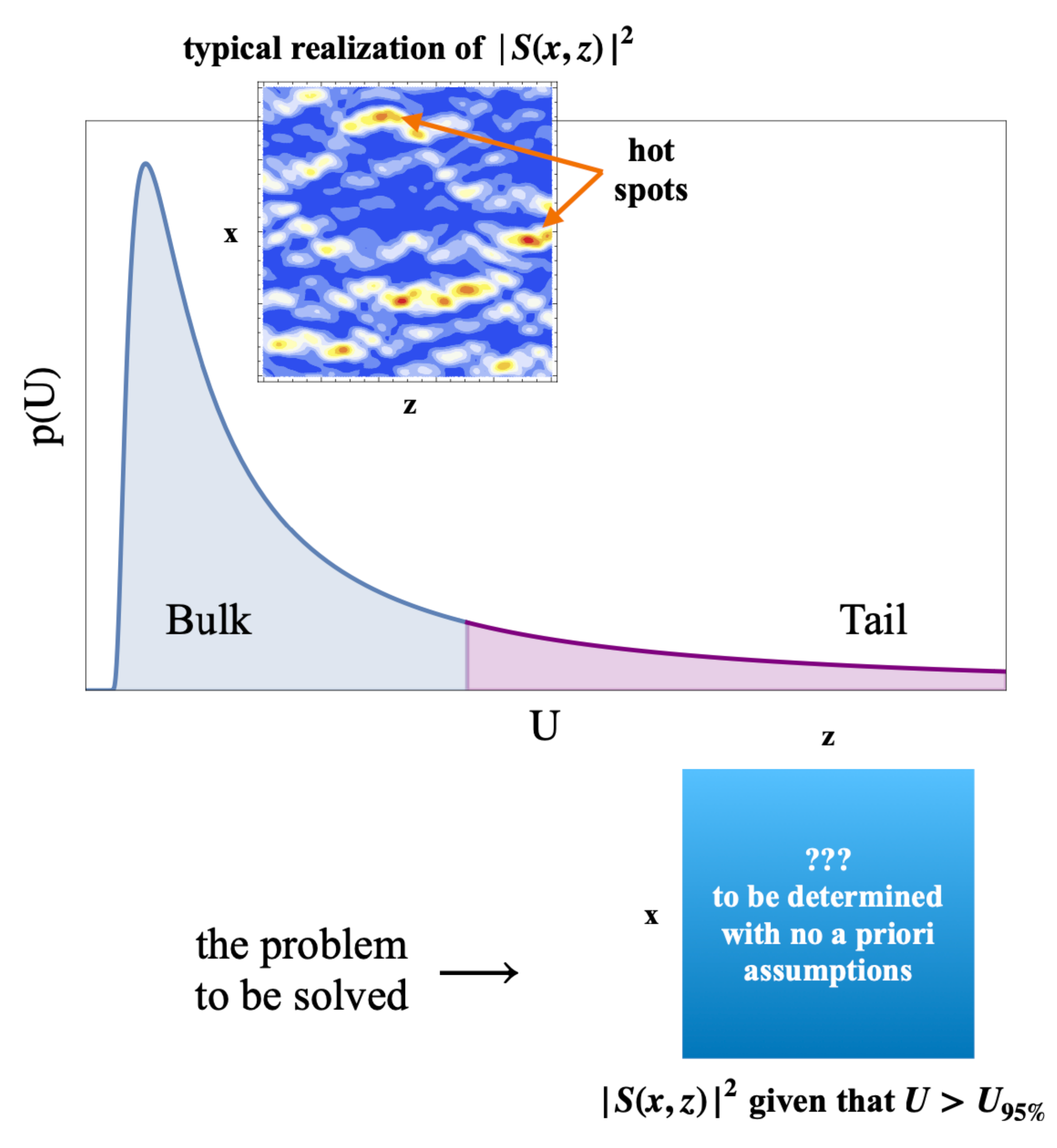}
\caption{\label{figure4}Illustration of the aim of Refs.~\onlinecite{Mounaix2023,Mounaix2024}. The Rose\textendash DuBois ansatz of hot spot fields in the tail of $p(U)$ (see Fig.~\ref{figure3}) is replaced with the problem to be solved: determine the realizations of $S(x,z)$ in the tail of $p(U)$ with no a priori assumptions.}
\end{figure}

This is the subject of the present paper. The aim is to explain as simply as possible the results recently obtained in Refs.~\onlinecite{Mounaix2023,Mounaix2024} concerning the validity of the Rose\textendash DuBois ansatz. The challenge was to determine the realizations of $S(x,z)$ in the tail of $p(U)$, without any a priori assumption \mbox{-- in} particular, without assuming (or ruling out) hot spot fields. Figure~\ref{figure4} summarizes this objective. Quite unexpectedly, the answer turns out to be sharply distinct from the hot spot of Fig.~\ref{figure3}, as can be observed in Figs.~\ref{figure10} and \ref{figure15}.

The paper is organized as follows. In Section~\ref{inst}, we outline the functional integral formalism and instanton analysis used in Ref.~\onlinecite{Mounaix2023} for determining the far upper tail of $p(U)$, the corresponding realizations of $S(x,z)$, and the value of $g_c(L)$, analytically. Section~\ref{test} is devoted to numerical results obtained in Ref.~\onlinecite{Mounaix2024} from a specific biased sampling procedure to probe the far upper tail of $p(U)$ and test the instanton analysis. In Section~\ref{trans}, we present some numerical results of Ref.~\onlinecite{Mounaix2023} for $U$ in the near upper tail of $p(U)$ (i.e., between the bulk and the far upper tail considered in Secs.~\ref{inst} and~\ref{test}). Taken together, these results sketch a complete picture of what the realizations of $S(x,z)$ look like as a function of $U$, from the bulk to the far upper tail of $p(U)$. The potential implications for laser-plasma interaction are discussed in Sec.~\ref{conse}. Finally, some perspectives are given in Sec.~\ref{persp}.
%
%
\section{The instanton analysis}\label{inst}
\subsection{The functional integral formalism}\label{thefif}
To begin with, we need to find an approach free of any prior assumptions about the realizations of $S(x,z)$, so that the results can be compared with those derived from the Rose\textendash DuBois ansatz. A good source of inspiration may be found in studies of stochastic fields of various kinds, initiated in the mid-1990s\cite{Giles1995,FKLM1996,SGMG2022,AMPV2022,GGS2015,FKLT2001,GM1996,BM1996} and building on the functional integral formalism developed in the mid-1970s\cite{Janssen1976,DeDominicis1976,DDP1978,Phythian1977,JP1979,Jensen1981}. In this respect, the paper by Falkovich {\it et al.} entitled "Instantons and Intermittency"\cite{FKLM1996} is particularly relevant, as evidenced by the opening sentence of the abstract: "We describe the method for finding the [...] tails of the probability distribution function for solutions of a stochastic differential equation [...]". The link to what we are looking for is clear: (i) finding the tail of $p(U)$ requires knowledge of the realizations of $S(x,z)$ that constitute it, and (ii) knowing the tail of $p(U)$ allows us to compute $g_c(L)$. The approach followed in Ref.~\onlinecite{Mounaix2023} was therefore to adapt and apply the method of Falkovich {\it et al.} in Ref.~\onlinecite{FKLM1996} to the stochastic amplifier~(\ref{withDeq}), as we will now explain.

For simplicity we will take $\psi(x,0)=1$ and we will assume that there is only one perpendicular direction $x$ ($d=1$) with periodic boundary conditions at $x=\pm\ell/2$ (i.e., we take for $\Gamma$ in Eq.~(\ref{withDeq}) the circle of length $\ell$). The starting point is to write $p(U)$ formally as the functional integral
\begin{equation}\label{generalactionint}
p(U)=\int_{\varphi(x,0)=1}
\delta(U-\vert\varphi(0,L)\vert^2)\, {\rm e}^{\mathcal{A}}\, 
\mathcal{D}^2\varphi\, \mathcal{D}^2\tilde{\varphi}\, \mathcal{D}^2 S,
\end{equation}
where the integration variables $\varphi$ and $\tilde{\varphi}$ are complex fields called the Martin-Siggia-Rose (MSR) conjugate fields\cite{MSR1973}, $S$ is the laser field, $\mathcal{D}^2\equiv\mathcal{D}{\rm Re}(\cdot)\mathcal{D}{\rm Im}(\cdot)$, and $\mathcal{A}\equiv\mathcal{A}(\varphi,\tilde{\varphi},S)$ plays the role of an action often referred to as the MSR action in the literature. In the case of Eq.~(\ref{withDeq}), $\mathcal{A}(\varphi,\tilde{\varphi},S)$ reads
\begin{widetext}
\begin{equation}\label{MSRactionwithD}
\mathcal{A}(\varphi,\tilde{\varphi},S)=
\frac{i}{2}\, \left\lbrack\int_{-\ell/2}^{\ell/2}\int_0^L \tilde{\varphi}^\ast(x,z) \left(D_{z,\, x^2}\varphi(x,z)
-g\vert S(x,z)\vert^2\varphi(x,z)\right) dz\, dx +c.\, c.\right\rbrack
-\int_{-\ell/2}^{\ell/2}\int_0^L S^\ast(x,z) (T_C^{-1}S)(x,z)\, dz\, dx,
\end{equation}
\end{widetext}
where $D_{z,\, x^2}=\partial_z -(i/2m)\partial^2_{x^2}$ and $T_C$ is the covariance operator of $S(x,z)$ defined by
\begin{equation}\label{covariance3D1}
(T_C f)(x,z)=\int_{-\ell/2}^{\ell/2}\int_0^L
C(x-x^\prime ,z-z^\prime)\, f(x^\prime ,z^\prime)\, dz^\prime\, dx^\prime ,
\end{equation}
where $C(x-x^\prime ,z-z^\prime)=\langle S(x,z)S(x^\prime ,z^\prime)^\ast \rangle$ is the autocorrelation function of $S(x,z)$ and $f(x,z)$ is a square-integrable function.

Note that in Ref.~\onlinecite{Mounaix2023}, $S(x,z)$ is not assumed to be homogeneous along $z$, only along $x$. In this case, one simply replaces $C(x-x^\prime ,z-z^\prime)$ with $C(x-x^\prime ,z,z^\prime)$, and $\langle\vert S(x,z)\vert^2\rangle =1$ with $L^{-1}\int_0^L \langle\vert S(x,z)\vert^2\rangle\, dz =1$, all other things being equal.
\subsection{The saddle-point equations and how to solve them}\label{spequs}
The point of writing $p(U)$ in the functional integral form~(\ref{generalactionint}) is that, in the large $\ln U$ limit, the tail of $p(U)$ is given by the saddle-point approximation of the right-hand side of Eq.~(\ref{generalactionint}) which, in some cases, can be computed explicitely. The saddle points that determine $p(U)$ in this limit are stationary points of $\mathcal{A}(\varphi,\tilde{\varphi},S)$, subject to the restriction $\vert\varphi(0,L)\vert^2=U$, which is imposed by the delta function in the functional integral~(\ref{generalactionint}). Following the usual procedure of Lagrange multipliers\cite{CH1989} to deal with the constraint, one obtains four stationarity conditions, or {\it saddle-point equations}. Three differential equations,
\begin{eqnarray}\label{insteqwithDphi2}
&&\left\lbrack D_{z,\, x^2}-g\vert S(x,z)\vert^2\right\rbrack\varphi(x,z)=0, \nonumber \\
&&\left\lbrack D_{z,\, x^2}+g\vert S(x,z)\vert^2\right\rbrack\tilde{\varphi}(x,z)=0, \\
&&\left\lbrack D_{z,\, x^2}^\ast +g\vert S(x,z)\vert^2\right\rbrack\tilde{\vartheta}(x,z)=0, \nonumber
\end{eqnarray}
where $D_{z,\, x^2}^\ast=\partial_z +(i/2m)\partial^2_{x^2}$, with boundary conditions
\begin{eqnarray}\label{boundarycond}
&&\varphi(x,0)=1, \nonumber \\
&&\tilde{\varphi}(x,L)=2i\lambda\varphi(0,L)\delta(x), \\
&&\tilde{\vartheta}(x,L) =2i\lambda\varphi^\ast(0,L) \delta(x) , \nonumber
\end{eqnarray}
where $\lambda$ is a Lagrange multiplier, and one integral equation,
\begin{equation}\label{insteqwithDS2}
\left\lbrack T_C\left(\tilde{\vartheta}\varphi +\tilde{\varphi}\varphi^\ast\right) S\right\rbrack (x,z)
=\frac{2i}{g}S(x,z).
\end{equation}
In Eqs.~(\ref{insteqwithDphi2}) to (\ref{insteqwithDS2}), $\tilde{\vartheta}\equiv\tilde{\varphi}^\ast$ and $\tilde{\varphi}$ are treated as independent fields.

Solving this coupled integro-differential system is beyond the scope of a this paper. Nevertheless, it may be interesting to explain how it can be done. The strategy adopted in Ref.~\onlinecite{Mounaix2023} follows a three-step roadmap. First, fix $S$ and solve (\ref{insteqwithDphi2}) (with boundary conditions (\ref{boundarycond})) for $\varphi$, $\tilde{\varphi}$, and $\tilde{\vartheta}$ formally as path integrals. This is the easy part, Eqs.~(\ref{insteqwithDphi2}) are Schr\"{o}dinger equations (with a complex potential) the path-integral solutions of which are well known\cite{FH1965,Schulman1981,Kleinert1995}. One obtains
\begin{eqnarray}\label{firststeproadmap}
&&\varphi(x,z)=\int_{-\ell/2}^{\ell/2}K(x,z;y,0)\, dy, \nonumber \\
&&\tilde{\varphi}(x,z)=2i\lambda\varphi(0,L)K(0,L;x,z)^\ast , \\
&&\tilde{\vartheta}(x,z)=2i\lambda\varphi^\ast(0,L)K(0,L;x,z) , \nonumber
\end{eqnarray}
where $K(x_2,z_2;x_1,z_1)$ is the Feynman-Kac propagator,
\begin{equation}\label{FKpropagator}
K(x_2,z_2;x_1,z_1)=\int_{x(z_1)=x_1}^{x(z_2)=x_2}{\rm e}^{\int_{z_1}^{z_2}
\left\lbrack\frac{im}{2}\dot{x}(\tau)^2+g\vert S(x(\tau),\tau)\vert^2\right\rbrack\, d\tau}
\mathcal{D}x,
\end{equation}
with $z_2>z_1$, where the path-integral is over all the continuous paths satisfying $x(z_1)=x_1$ and $x(z_2)=x_2$. Then (second step), inject the expressions~(\ref{firststeproadmap}) onto the left-hand side of Eq.~(\ref{insteqwithDS2}) to get the equation for $S(x,z)$,
\begin{equation}\label{insteqwithDS3}
\varphi^\ast(0,L)\, G_1(x,z)+\varphi(0,L)\, G_2(x,z)=\frac{1}{\lambda g}\, S(x,z),
\end{equation}
with
\begin{eqnarray}\label{integralG1}
G_1(x,z)=&&\int_0^L\int_\Lambda\int_\Lambda
K(0,L;x^\prime,z^\prime)K(x^\prime,z^\prime;\xi,0) \nonumber \\
&&\times C(x-x^\prime,z-z^\prime)\, S(x^\prime,z^\prime)
\, dx^\prime\, d\xi\, dz^\prime ,
\end{eqnarray}
and
\begin{eqnarray}\label{integralG2}
G_2(x,z)=&&\int_0^L\int_\Lambda\int_\Lambda
K(0,L;x^\prime,z^\prime)^\ast K(x^\prime,z^\prime;\xi,0)^\ast \nonumber \\
&&\times C(x-x^\prime,z-z^\prime)\, S(x^\prime,z^\prime)
\, dx^\prime\, d\xi\, dz^\prime .
\end{eqnarray}
It can be seen in Eq.~(\ref{FKpropagator}) that the Feynman-Kac propagators on the right-hand side of Eqs.~(\ref{integralG1}) and (\ref{integralG2}) depend on $S$ in a highly non-trivial way. Consequently, Eq.~(\ref{insteqwithDS3}) is a complicated functional equation for $S(x,z)$ that must be solved. This is the hard part. The solutions together with the corresponding fields in Eqs.~(\ref{firststeproadmap}) are the stationary points of the MSR action with the restriction $\vert\varphi(0,L)\vert^2=U$, among which it remains to extract the saddle points needed to determine the tail of $p(U)$ (third and last step).

In the large $\ln U$ limit, Eq.~(\ref{insteqwithDS3}) can be solved explicitly for at least one class of $S(x,z)$. In that case, the realizations of $S(x,z)$ in the tail of $p(U)$, the expression of the tail, and the corresponding value of $g_c(L)$ can all be obtained. The rest of the section is devoted to presenting the results. The interested reader will find the details of the calculations in Ref.~\onlinecite{Mounaix2023}.
\subsection{The instanton solutions}\label{theinstsol}
First of all, we give some definitions that will be needed in the following.
\begin{description}
\item[def. 1]
$T\lbrack x(\cdot)\rbrack$ is the covariance operator defined by
\begin{equation}\label{pathcovar}
(T\lbrack x(\cdot)\rbrack f)(z)=\int_0^L
C(x(z)-x(z^\prime) ,z-z^\prime)\, f(z^\prime)\, dz^\prime ,
\end{equation}
where $x(\cdot)$ is a continuous path arriving at $x(L)=0$ (where $U$ is computed) and $f(z)$ is a square-integrable function;
\item[def. 2]
$\mu_1\lbrack x(\cdot)\rbrack$ is the largest eigenvalue of $T\lbrack x(\cdot)\rbrack$;
\item[def. 3]
$x_{\rm inst}(\cdot)$ is a (not necessarily continuous) path arriving at $x_{\rm inst}(L)=0$ and maximizing $\mu_1\lbrack x(\cdot)\rbrack$.
\end{description}
Since $C$ in Eq.~(\ref{pathcovar}) is the correlation function of $S(x(z),z)$, it is a positive definite kernel, and all the eigenvalues of $T\lbrack x(\cdot)\rbrack$ are real and positive. In def. 3, the subscript {\it inst} stands for {\it instanton} whose meaning is given below. The class of $S(x,z)$ considered in Ref.~\onlinecite{Mounaix2023} is defined by the following three points:
\begin{description}
\item[point 1]
there is a finite number of $x_{\rm inst}(\cdot)$s;
\item[point 2]
all the $x_{\rm inst}(\cdot)$s are continuous paths;
\item[point 3]
the fundamental eigenspaces of $T\lbrack x_{\rm inst}(\cdot)\rbrack$ for different $x_{\rm inst}(\cdot)$s (if any) are essentially disjoint (i.e., their intersection reduces to the zero vector).
\end{description}
Although restricted, this class of $S(x,z)$ is quite large as it includes most of the RPP laser fields encountered in laser-plasma interaction.

The saddle-points that determine $p(U)$ in the asymptotic regime $\ln U\gg 1$ are the saddle-points with the largest value of $\mathcal{A}(\varphi,\tilde{\varphi},S)$. These highest saddle-points are called {\it leading instantons} or, more simply, {\it instantons}. An important consequence of the point 3 above is the existence of a one-to-one correspondence between the paths $x_{\rm inst}(\cdot)$ and the instantons. Each instanton can be labeled by the $x_{\rm inst}(\cdot)$ it is associated with, and for a given $x_{\rm inst}(\cdot)$, the $S$-component of the corresponding instanton can be determined explicitly, in the large $\ln U$ limit. Assuming for simplicity that $\mu_1\lbrack x_{\rm inst}(\cdot)\rbrack$ is not degenerate (the degenerate case is a little more technical with not substantially different results) and writing $\mu_1\lbrack x_{\rm inst}(\cdot)\rbrack\equiv\mu_{\rm max}(L)$ to make explicit the fact that $\mu_1\lbrack x_{\rm inst}(\cdot)\rbrack$ depends on $L$ and not on the specific $x_{\rm inst}(\cdot)$ (if there is more than one), one finds that\cite{Mounaix2023}
\begin{equation}\label{notdegesingfil}
S_{\rm inst}^{x_{\rm inst}(\cdot)}(x,z)=\frac{c_1}{\sqrt{\mu_{\rm max}(L)}}
\, \int_0^L C(x-x_{\rm inst}(z^\prime),z-z^\prime)
\, \phi_1(z^\prime)\, dz^\prime ,
\end{equation}
where $S_{\rm inst}^{x_{\rm inst}(\cdot)}(x,z)$ denotes the $S$-component of the instanton associated with the path $x_{\rm inst}(\cdot)$, $\phi_1(z)$ is the normalized fundamental eigenmode of $T\lbrack x_{\rm inst}(\cdot)\rbrack$, and $c_1$ is a complex Gaussian random variable with $\langle c_1\rangle=\langle c_1^2\rangle=0$ and $\langle\vert c_1\vert^2\rangle=1$. The asymptotic regime $\ln U\gg 1$ then coincides with the limit of high-intensity instantons $\vert c_1\vert^2\gg 1$.

The expression of $S_{\rm inst}^{x_{\rm inst}(\cdot)}(x,z)$ in Eq.~(\ref{notdegesingfil}) provides a good illustration of the general property of instantons to be quasi-deterministic, i.e., much less random than $S(x,z)$ itself. While the number of random variables characterizing $S(x,z)$ is equal to the number of RPP elements in the laser cross section ($\gtrsim 10^4$), it takes only one (complex) random amplitude, $c_1$, to entirely characterize the instanton~(\ref{notdegesingfil}). In the degenerate case, the number of random variables characterizing the instanton is equal to the degeneracy of $\mu_1\lbrack x_{\rm inst}(\cdot)\rbrack$ ($\ll 10^4$).
\begin{figure*}[!t]
\includegraphics[width=0.75\linewidth]{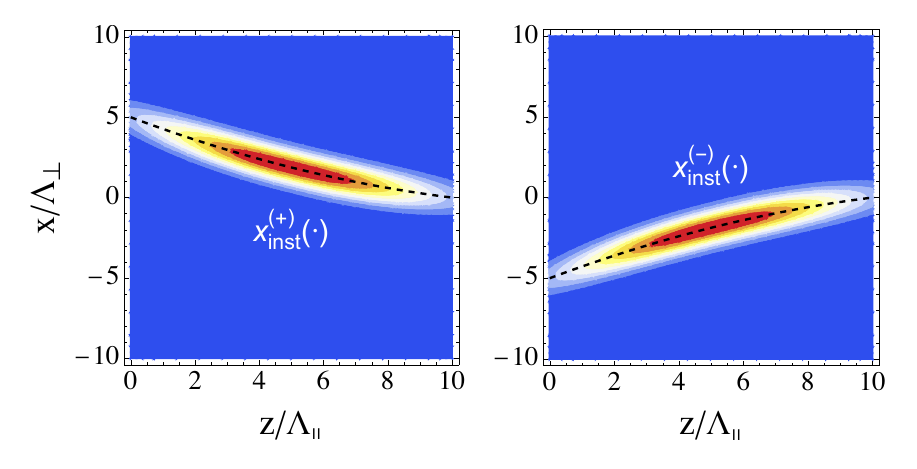}
\caption{\label{figure5}Artist's rendering of $\vert S_{\rm inst}^{x_{\rm inst}(\cdot)}(x,z)\vert^2$ for two non-degenerate single-filament instantons in the symmetric case $C(-x,z)=C(x,z)$ with two different paths $x^{\pm}_{\rm inst}(\cdot)$ maximizing $\mu_1\lbrack x(\cdot)\rbrack$ (dashed lines). Note that $C(-x,z)=C(x,z)$ implies $\mu_1\lbrack -x(\cdot)\rbrack =\mu_1\lbrack x(\cdot)\rbrack$ and $x^{\pm}_{\rm inst}(\cdot)$ are symmetric with respect to $x=0$. (See defs. 1 to 3 and points 1 to 3 in the text.)}
\end{figure*}

It may be instructive to compare the expression (\ref{notdegesingfil}) with its counterpart for a high-intensity hot spot, say at the origin of coordinates\cite{RD1993,Adler1981,Mounaix2015,Mounaix2019},
\begin{equation}\label{theorhotspot}
S_{HS}(x,z)=c_{HS}\, C(x,z),
\end{equation}
where HS stands for {\it hot spot} and $c_{HS}=S(0,0)$ knowning that $\vert S(x,z)\vert^2$ has a high local maximum at the origin\cite{RD1993,Garnier1999}. Equations~(\ref{notdegesingfil}) and (\ref{theorhotspot}) are both the product of a random amplitude (resp. $c_1$ and $c_{HS}$) and a deterministic (non-random) profile. However, they differ significantly in how the resulting structure is distributed. From Eq.~(\ref{theorhotspot}), we see that the hot spot remains localized, with $C(x,z)$ confined to a correlation volume centered at the origin. In contrast, the convolution on the right-hand side of Eq.~(\ref{notdegesingfil}) spreads the instanton along the entire path $x_{\rm inst}(\cdot)$, resulting in an elongated, filament-like structure. These instantons are referred to as {\it filamentary instantons} in Ref.~\onlinecite{Mounaix2023}, and for the class of $S(x,z)$ considered, they are {\it single-filament} instantons.

Figure~\ref{figure5} shows realistic artist's renderings of $\vert S_{\rm inst}^{x_{\rm inst}(\cdot)}(x,z)\vert^2$ for two single-filament instantons under the assumption that $\mu_1\lbrack x(\cdot)\rbrack$ admits two $x_{\rm inst}(\cdot)$s symmetric with respect to $x=0$ (denoted by $x^{\pm}_{\rm inst}(\cdot)$ in the figure). A fully computed example of a single-filament instanton will be provided in Sec.~\ref{test}. In the meantime, Fig.~\ref{figure5} offers a faithful visual representation of what such instantons typically look like. Specifically, the characteristic instanton length and width are $\sim L$ and $\sim\Lambda_{\perp}$, respectively. For $L\gg\Lambda_{\perp}$, the instanton is indeed confined within a narrow tube -- or filament -- along the path $x_{\rm inst}(\cdot)$ it is associated with. This structure persists in the degenerate case with a quasi-deterministic, instead of non-random, instanton profile. By contrast, a high-intensity hot spot would appear as a disk of unit diameter.

\begin{figure}[!b]
\includegraphics[width=0.75\linewidth]{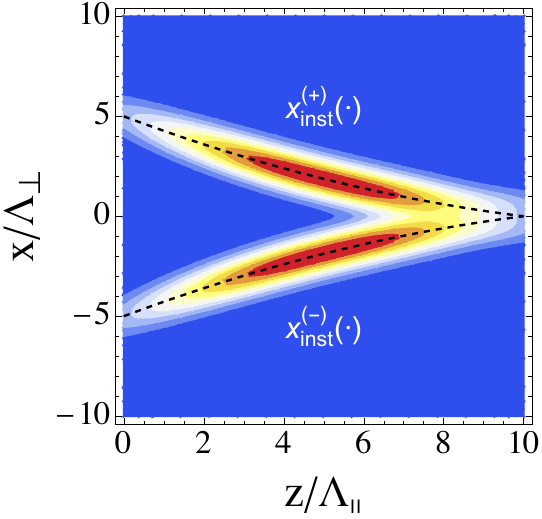}
\caption{\label{figure6}Artist's rendering of $\vert S_{\rm inst}(x,z)\vert^2$ for one non-degenerate two-filament instanton for the same case as in Fig.~\ref{figure5}, but with coincident instead of essentially disjoint fundamental eigenspaces of $T\lbrack x_{\rm inst}^{\pm}(\cdot)\rbrack$.}
\end{figure}

Without going into detail, it should be noted that single-filament instantons are not the only possible instantons. The number of filaments in an instanton depends on both the number of $x_{\rm inst}(\cdot)$s and the intersections of the fundamental eigenspaces of $T\lbrack x_{\rm inst}(\cdot)\rbrack$ for the different $x_{\rm inst}(\cdot)$s. If all these eigenspaces are essentially disjoint (point 3), then all the instantons are single-filament instantons. But multi-filament instantons do become possible if point 3 is lifted. Consider for instance the case illustrated in Fig.~\ref{figure5} of two non-degenerate single-filament instantons along $x_{\rm inst}^{\pm}(\cdot)$. Since the instantons are not degenerate, the fundamental eigenspaces of $T\lbrack x_{\rm inst}^{\pm}(\cdot)\rbrack$ are one-dimensional, and if they are not essentially disjoint, then they coincide. In this case, the fundamental eigenmode of $T\lbrack x_{\rm inst}^{+}(\cdot)\rbrack$ is also the fundamental eigenmode of $T\lbrack x_{\rm inst}^{-}(\cdot)\rbrack$ and vice versa, and the two non-degenerate single-filament instantons of Fig.~\ref{figure5} merge into one non-degenerate two-filament instanton, as shown in Fig.~\ref{figure6}.

Note that in Figs.~\ref{figure5} and \ref{figure6}, the dashed lines representing the $x_{\rm inst}(\cdot)$s have been carefully drawn along the ridge paths where $\vert S_{\rm inst}(x,z)\vert^2$ reaches a global maximum for each $z$. This graphical choice is not merely for visual clarity. As shown in Ref.~\onlinecite{Mounaix2023}, the paths maximizing $\mu_1\lbrack x(\cdot)\rbrack$ {\it are} the ridge paths of $\vert S_{\rm inst}(x,z)\vert^2$. Physically, these paths are the trajectories along which the amplification of $\varphi_{\rm inst}(x,z)$ is maximum in the large $U$ limit. This is why they play a central role in the instanton solution of Eq.~(\ref{withDeq}). Remarkably, as paths maximizing $\mu_1\lbrack x(\cdot)\rbrack$ (which is not random), the $x_{\rm inst}(\cdot)$s are non-random paths. Even in the degenerate case where the instanton profile itself is random, the ridge paths remain non-random. They constitute the deterministic skeleton of the filamentary instanton structure.

Before presenting the results of the instanton analysis, it may be useful to briefly discuss the instanton solution in the small length limit defined by $L\ll\Lambda_{\parallel}$. This regime becomes physically relevant whenever plasma inhomogeneities\cite{Kruer1988,Rosenbluth1972,PLP1973,Liu1976} are strong enough to make $L$ smaller than one hot spot length\cite{TMP1997,FWFTP2022}. Writing $\varepsilon_{\parallel}\equiv L/\Lambda_{\parallel}$ and $\varepsilon_{\perp}\equiv \vert x\vert/\Lambda_{\perp}$, one finds that in the small length ($\varepsilon_{\parallel}\ll 1$) and near axis ($\varepsilon_{\perp}\ll 1$) limits, the right-hand side of Eq.~(\ref{notdegesingfil}) reduces to
\begin{equation}\label{smallsystlim}
S_{\rm inst}^{x_{\rm inst}(\cdot)}(x,z)=c_1\left\lbrack C\left(x,z-\frac{L}{2}\right)
+o(\varepsilon_{\parallel}^2)
+o(\varepsilon_{\perp}^2)\right\rbrack ,
\end{equation}
from which it follows that the profile of $S_{\rm inst}^{x_{\rm inst}(\cdot)}(x,z)$ coincides with the hot spot profile of $S_{HS}(x,z-L/2)$ in Eq.~(\ref{theorhotspot}), up to second order in $\varepsilon_{\parallel ,\, \perp}\ll 1$. Note however that $c_{HS}$ in Eq.~(\ref{theorhotspot}) and $c_1$ in Eq.~(\ref{smallsystlim}) are distributed differently. Writing $p_{HS}(u)$ and $p_{{\rm inst}}(u)$ the probability distributions of $u=\vert c_{HS}\vert^2$ and $u=\vert c_1\vert^2$ respectively, one has\cite{Garnier1999} $p_{HS}(u)\sim u^{(d+1)/2}\exp(-u)$ ($u\gg 1$) while $p_{{\rm inst}}(u)=\exp(-u)$. Whether or not there is an algebraic term in front of the exponential comes from the fact that the two fields $S_{HS}(x,z-L/2)$ and $S_{\rm inst}(x,z)$ are not conditioned in the same way. The former is conditioned on $\vert S(0,L/2)\vert^2$ being a high local maximum of $\vert S(x,z)\vert^2$, while the latter is conditioned on $\vert\psi(0,L\vert^2$ being large.

The computation of the critical coupling $g_c(L)$ from the instanton (\ref{smallsystlim}) is similar to the one from the hot spot (\ref{theorhotspot}) where $p_{HS}(u)$ is replaced with $p_{\rm inst}(u)$. The absence of an algebraic term in front of the exponential in $p_{\rm inst}(u)$ does not affect the onset of the divergence of $\langle U\rangle$ and, in the limit $L\ll\Lambda_{\parallel}$, both Rose\textendash DuBois ansatz and instanton analysis give the same value of $g_c(L)$, up to second order in $\varepsilon_{\parallel}$. It can be shown\cite{Mounaix2001} that the results differ at higher order and that the difference is caused by random fluctuations of $S(x,z)$ around the theoretical hot spot profile. This is in contrast to the crucial assumption of the Rose\textendash DuBois ansatz that the higher the hot spot intensity, the more negligible the effects of hot spot profile fluctuations. According to this idea, the divergence of $\langle U\rangle$ -- and therefore the value of $g_c(L)$ -- should not depend on profile fluctuations. The fact that this is not the case confirms that the divergence of $\langle U\rangle$ is not caused by high-intensity hot spots but by another type of structures, close to (but different from) hot spots in the limit $L\ll\Lambda_{\parallel}$. After that digression, we now turn to the results of the instanton analysis.
\subsection{The results of the instanton analysis}\label{theresultsofia}
The saddle-point approximation of Eq.~(\ref{generalactionint}) when $\ln U\gg 1$ is based on the same idea as the standard Laplace's method\cite{BO1978} according to which, in this limit, fluctuations orthogonal to the instantons are small and can be integrated out as standard Gaussian fluctuations, yielding
\begin{equation}\label{asymlogofpofu}
\ln p(U)\sim\ln\sum_{x_{\rm inst}(\cdot)}\left\langle
\delta(U-\vert\varphi_{\rm inst}^{x_{\rm inst}(\cdot)}(0,L)\vert^2)
\right\rangle_{S_{\rm inst}^{x_{\rm inst}(\cdot)}}\, .
\end{equation}
Here, $\langle\cdot\rangle_{S_{\rm inst}}$ denotes the average over the realizations of $S_{\rm inst}$. Building on this expression and the results discussed in the previous section, we are now in a position to address the questions raised in Sec.~\ref{intro}. Specifically: (i) the realizations of $S(x,z)$ in the tail of $p(U)$, (ii) the analytical expression for this tail, and (iii) the corresponding value of the critical coupling $g_c(L)$.

\smallskip
{\it (i) The realizations of $S(x,z)$ given that $\ln U\gg 1$}. Each term in the sum on the right-hand side of Eq.~(\ref{asymlogofpofu}) is similar to the exact expression, $p(U)=\langle\delta(U-\vert\psi(0,L)\vert^2)\rangle$, for the probability of $U$ in problem (\ref{withDeq}) where $S$ is replaced by $S_{\rm inst}^{x_{\rm inst}(\cdot)}$. This means that, in the large $\ln U$ limit, $p(U)$ can be determined by considering only those realizations of $S(x,z)$ that are instanton realizations. The contribution from other realizations is negligible with respect to the leading terms in Eq.~(\ref{asymlogofpofu}). In other words, the dominant realizations of $S(x,z)$ in the far upper tail of $p(U)$ are the filamentary instantons described in Sec.~\ref{theinstsol}.

\smallskip
{\it (ii) The tail of $p(U)$}. From the first Eq.~(\ref{firststeproadmap}) with $S$ given by Eq.~(\ref{notdegesingfil}), it is possible to compute the right-hand side of Eq.~(\ref{asymlogofpofu}) explicitely. One finds\cite{Mounaix2023}
\begin{equation}\label{logfartailofpofu}
\ln p(U)\sim -\left(1+\frac{1}{2g\mu_{\rm max}(L)}\right)\, \ln U\ \ \ \ \ (\ln U\gg 1),
\end{equation}
from which it follows that the far upper tail of $p(U)$ takes the form of a leading algebraic tail $\propto U^{-\zeta}$ with exponent $\zeta=1+1/2g\mu_{\rm max}(L)$, modulated by a slow varying amplitude $f(U)$ (slower than algebraic),
\begin{equation}\label{fartailofpofu}
p(U)\sim f(U)U^{-\zeta}\ \ \ \ \ (\ln U\gg 1).
\end{equation}
Determining $f(U)$ requires going beyond the leading-order terms in Eq.~(\ref{logfartailofpofu}), which will not be necessary for our present purposes.

\smallskip
{\it (iii) The critical coupling $g_c(L)$}. Injecting this result into $\langle U\rangle=\int^{+\infty}Up(U)\, dU$, one finds that the integral diverges for all $g>1/2\mu_{\rm max}(L)$ (and converges for all $g<1/2\mu_{\rm max}(L)$). Therefore, the value of the critical coupling given by the instanton analysis is 
\begin{equation}\label{critcoupling}
g_c(L)=\frac{1}{2\mu_{\rm max}(L)}.
\end{equation}

The next step is to test the results of the instanton analysis presented in this section by comparing them either with analytical predictions from alternative approaches (when available) or with numerical simulations. This was the focus of Ref.~\onlinecite{Mounaix2024}, whose main results are presented in the next section.
\begin{figure*}[!t]
\includegraphics[width=0.75\linewidth]{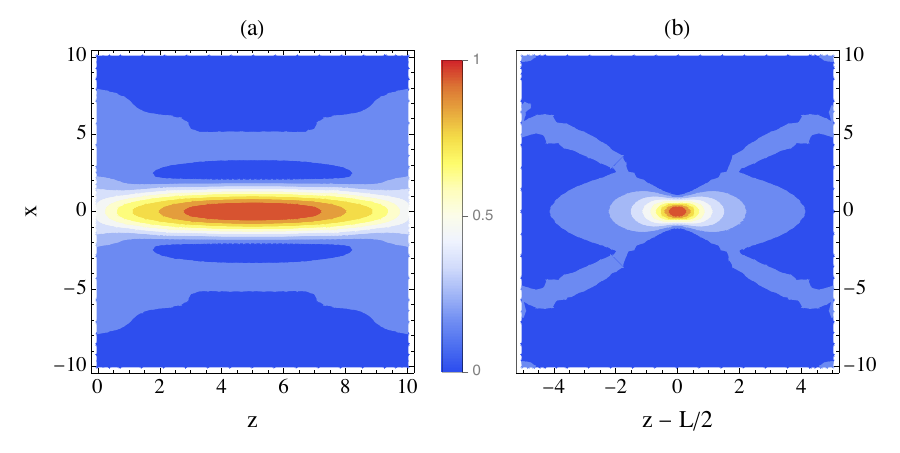}
\caption{\label{figure7}Contour plots of (a) $\vert S_{\rm inst}(x,z)\vert^2$ normalized to $\max\vert S_{\rm inst}(x,z)\vert^2=1$ and (b) the hot spot profile $\vert C(x,z)\vert^2$, for $S(x,z)$ given by Eq.~(\ref{solSnum}) with $\ell=20$ and $L=10$.}
\end{figure*}
%
%
%
\section{Testing the instanton analysis}\label{test}
\subsection{Cross-verifying analytical predictions for $\bm{g_c(L)}$}\label{cvtpcritical}
To test the validity of the critical coupling~(\ref{critcoupling}), we can rely on the exact analytical result from Ref.~\onlinecite{MCL2006}, which was derived using a completely different approach. In that work, the question of whether $\langle U\rangle$ diverges is addressed through a distributional formulation of $\psi(0,L)$ combined with the use of the Paley-Wiener theorem to control its growth, thereby allowing one to assess the conditions under which $\langle U\rangle$ remains finite or diverges. It is important to note that, in this approach, there is no need to know the tail of $p(U)$; therefore, an instanton analysis of the problem is not required. The trade-off, however, is that this method does not provide information about the specific realizations of $S(x,z)$ that cause the divergence. That said, the value of $g_c(L)$ in Eq.~(\ref{critcoupling}) coincides with the one in Ref.~\onlinecite{MCL2006}, which provides strong evidence for the robustness of Eq.~(\ref{critcoupling}), as it arises from at least two entirely different theoretical methods.
\subsection{Numerical results in the far upper tail of $\bm{p(U)}$}\label{numericsfartail}
The current absence of an alternative analytical derivation for both the tail of $p(U)$ and the corresponding realizations of $S(x,z)$, apart from the instanton analysis, makes it essential to test the latter's predictions numerically. The Gaussian field $S$ considered in both Ref.~\onlinecite{Mounaix2024} and the section of Ref.~\onlinecite{Mounaix2023} devoted to numerical results is
\begin{equation}\label{solSnum}
S(x,z)=\sum_{n=-50}^{50}s_{n}\sqrt{\varsigma_n}\, 
\exp\, i\left\lbrack\frac{2\pi n}{\ell} x+\left(\frac{2\pi n}{\ell}\right)^2\frac{z}{2}\right\rbrack ,
\end{equation}
where the $s_n$s are complex Gaussian random variables with $\langle s_n\rangle =\langle s_n s_m\rangle =0$ and $\langle s_n s_m^\ast\rangle =\delta_{nm}$, and the spectral density $\varsigma_n$ is normalized to $\sum_{n=-50}^{50}\varsigma_n=1$. Equation~(\ref{solSnum}) is reminiscent of models of spatially smoothed laser beams\ \cite{RD1993}, where $S$ is a solution to the paraxial wave equation
\begin{equation}\label{eqSnum}
\partial_z S(x,z)+\frac{i}{2}\partial^2_{x^2} S(x,z)=0,
\end{equation}
with $S(x,0)=\sum_{n=-50}^{50} s_n\sqrt{\varsigma_n}\, \exp(2i\pi nx/\ell)$. To ensure that the space average $\ell^{-1}\int_{\Lambda}S(x,z)\, dx$ is zero for all $z$ and every realization of $S$, as expected for the electric field of a smoothed laser beam, the mode at $n=0$ is excluded by taking $\varsigma_{0}=0$. Here we show the results for the Gaussian spectrum
\begin{equation}\label{gaussspectrnum}
\varsigma_{n\ne 0}\propto\exp\left\lbrack -\left(\frac{\pi n}{\ell}\right)^2\right\rbrack .
\end{equation}
(Other widely used spectra, like top-hat and Cauchy spectra, give similar results.)

As explained in Ref.~\onlinecite{Mounaix2023}, there is only one $x_{\rm inst}(\cdot)$ in this case, namely $x_{\rm inst}(\cdot)\equiv 0$, and this unique path supports a non-degenerate, single-filament instanton $S_{\rm inst}$ of the form (\ref{notdegesingfil}). Figure~\ref{figure7} shows the contour plots of the profile of $\vert S_{\rm inst}\vert^2$ and the hot spot profile $\vert C\vert^2$ for $L=10$ and $\ell=20$. The fully computed single-filament instanton in Fig.~\ref{figure7}(a) clearly exhibits all the features discussed in Sec.~\ref{theinstsol} and represented in the artist's rendering of Fig.~\ref{figure5}.

To test the instanton analysis predictions numerically, it is crucial to have a good sampling of the realizations of $S(x,z)$ for which $\ln U\gg 1$. Unfortunately, such realizations are extremely rare events, far beyond the reach of any direct sampling with a reasonable sample size. For instance, considering the same Gaussian field $S$ and parameters as in Ref.~\onlinecite{Mounaix2024}, and working in the simple diffraction-free limit, $m^{-1}=0$, where $U$ can be computed explicitly, one can verify that $p(\ln U\ge 10^3)=O(10^{-100})$. It is thus clearly unrealistic to expect that the results of the instanton analysis can be tested by direct numerical simulations. To gain access to the far upper tail of $p(U)$ and check the validity of the instanton analysis, a specific approach is absolutely necessary.

There exist several standard general methods for sampling rare and extreme events that could, in principle, be employed -- such as importance sampling algorithms\cite{HM1956,Hartmann2014,HDMRS2018,HMS2019} or filtering techniques\cite{AB2001,PBZS2015} (for massive data sets). However, all of these approaches are computationally intensive, both in terms of processing time and resource requirements. To the best of the author's knowledge, none of these methods have yet been applied to the study of stochastic amplifiers in the regime of large amplification. Adapting these techniques to the general context of stochastic amplification -- assuming such an adaptation is even feasible, which is not entirely obvious -- remains a substantial and yet unresolved challenge.

Fortunately, it turns out that for $S(x,z)$ specified in Eq.~(\ref{solSnum}) -- and more generally, for any $S$ that admits a single non-degenerate instanton (whether single- or multi-filament) -- it is possible to devise a tailored, much simpler biased sampling procedure that provides access to the far upper tail of $p(U)$. Such a procedure has been developed and implemented in Ref.~\onlinecite{Mounaix2024}. We will now present its main results.
\begin{figure}[!b]
\includegraphics[width=1.0\linewidth]{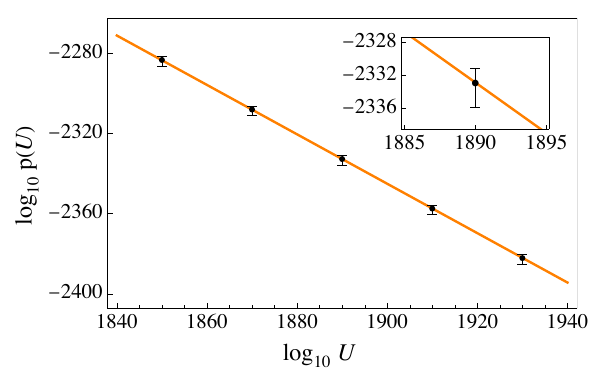}
\caption{\label{figure8}Numerical computation of $\log_{10}p(U)$ for $\log_{10}U$ between $1850$ and $1930$ by steps of $20$ (black dots). Plot of $\log_{10}p(U)=-\zeta\log_{10}U+{\rm Const}$ with $\zeta=1.22989$, as predicted by Eq.~(\ref{logfartailofpofu}), and ${\rm Const}=-8.41526$ adjusted to get the best fit to numerical data (solid line). Inset: enlargement near $\log_{10}U=1890$. (Fig. 6 of Ref.~\onlinecite{Mounaix2024}. Copyright 2024 IOP Publishing Ltd)}
\end{figure}

For each realization of $S(x,z)$ on a cylinder of length $L=10$ and circumference $\ell=20$, the equation~(\ref{withDeq}) has been solved numerically for $m=0.7$ and $g=0.5$. With these parameters, one has $\mu_{\rm max}(L)=4.34984$, $g_c(L)=0.11495$, yielding $g/g_c(L)\simeq 4.35>1$, and $\zeta=(1+g_c(L)/g)=1.22989$.

\begin{figure}[!b]
\includegraphics[width=1.0\linewidth]{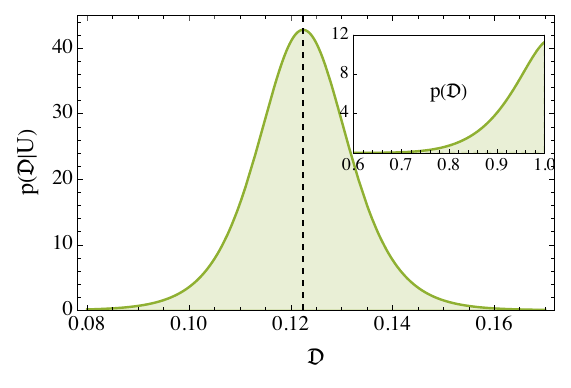}
\caption{\label{figure9}Distribution of $\mathfrak{D}$ conditioned on the value of $U$, $p(\mathfrak{D}\vert U)$, for fixed $\log_{10}U=1890$. The median of $p(\mathfrak{D}\vert U)$, here at $\mathfrak{D}=0.122457$, is indicated by a dashed vertical line. Inset: unconditioned distribution $p(\mathfrak{D})$. (Fig. 8 of Ref.~\onlinecite{Mounaix2024}. Copyright 2024 IOP Publishing Ltd)}
\end{figure}

\begin{figure*}[!t]
\includegraphics[width=0.75\linewidth]{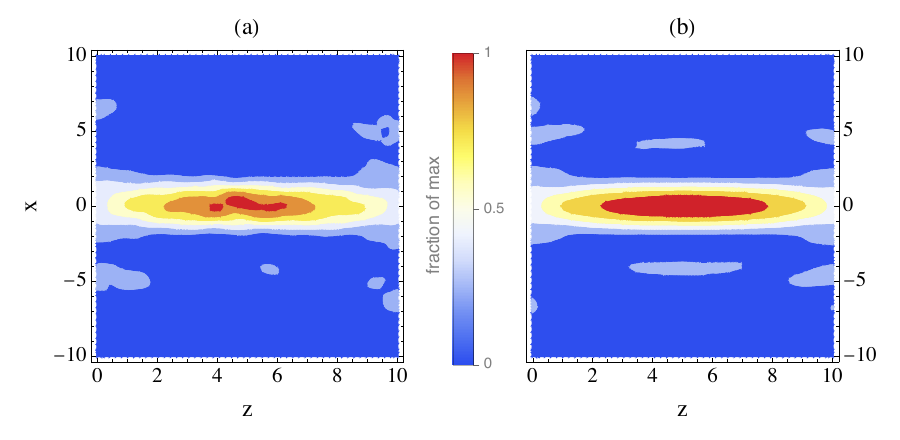}
\caption{\label{figure10}Contour plots of (a) $\vert S(x,z)\vert^2$ for a typical realization with $\log_{10}U=1890$ and (b) the sample mean of $\vert\hat{S}(x,z)\vert^2$ for $10^2$ realizations with $\log_{10}U=1890$. (Fig. 9 of Ref.~\onlinecite{Mounaix2024}. Copyright 2024 IOP Publishing Ltd)}
\end{figure*}

Figure~\ref{figure8} shows the numerical test of the analytical expression~(\ref{logfartailofpofu}) for the far upper tail of $p(U)$. Black dots are the numerical results and the solid line is the analytical result. We observe an almost perfect alignment of the numerical data along a straight line with slope $-\zeta$, which validates the algebraic tail (\ref{logfartailofpofu}) numerically, in the considered range of $\log_{10}U$. Note the extremely small values of $p(U)<10^{-2270}$, indeed unattainable without employing a tailored biased sampling procedure. The inset shows an enlarged view of the region where the realizations of $S(x,z)$ have been analysed in Ref.~\onlinecite{Mounaix2024}, with selected results presented in Figs.~\ref{figure9} and \ref{figure10}.

Figure~\ref{figure9} shows two distributions of the $L^2$-distance $\mathfrak{D}$ between the profile of $S(x,z)$ and its component along the instanton, with $\mathfrak{D}$ lying between $0$ and $1$ (see Ref.~\onlinecite{Mounaix2024} for details). $\mathfrak{D}=1$ means that the realization of $S(x,z)$ is orthogonal to the instanton, while $\mathfrak{D}=0$ means that $S(x,z)$ is an instanton realization. The inset shows the distribution of $\mathfrak{D}$ when no constraint is imposed on the value of $U$. It can be seen that the values of $\mathfrak{D}$ are concentrated near $1$, typically between $0.8$ and $1$, which is consistent with the absence of instantons in typical realizations of $S(x,z)$, as mentioned in Sec.~\ref{lhs}. The main panel of Fig.~\ref{figure9}, by contrast, shows the distribution of $\mathfrak{D}$ conditioned on the value of $U$, $p(\mathfrak{D}\vert U)$, for $\log_{10}U=1890$. With $\mathfrak{D}$ between $0.08$ and $0.16$, the corresponding realizations of $S(x,z)$ are visibly much closer to the instanton (in the $L^2$ sense) than typical realizations.

This proximity between the realizations of $S(x,z)$ in the far upper tail of $p(U)$ and the instanton becomes clearly apparent when directly plotting the profile of $S(x,z)$. Figure~\ref{figure10}(a) shows a contour plot of $\vert S(x,z)\vert^2$ for a typical realization, again with the same value of $\log_{10}U=1890$. As can be seen, it is the superposition of an elongated cigar-shaped component along $x=0$ and fluctuations, or noise, of comparatively small amplitude. The presence of noise results from $\ln U$ being large, though not asymptotically so ($\sim 5\ 10^3$). An effective way to suppress the noise and reveal the underlying structure is to average the profiles of a sufficiently large number of realizations, all with the same value of $U$ in the far upper tail of $p(U)$. Figure~\ref{figure10}(b) shows the sample mean of $\vert\hat{S}(x,z)\vert^2$, where $\hat{S}=S/\| S\|_2$, $\|\cdot\|_2$ denoting the $L^2$-norm, from $10^2$ different realizations with $\log_{10}U=1890$.

The close resemblance between Figs.~\ref{figure7}(a) and \ref{figure10}(b) is obvious and the dominant cigar-shaped component of $\vert S(x,z)\vert^2$ along $x=0$ visible in Fig.~\ref{figure10}(a) can clearly be identified as the theoretical instanton profile. This result, together with the small values of $\mathfrak{D}$ observed in Fig.~\ref{figure9}, show that the realizations of $S(x,z)$ in the far upper tail of $p(U)$ (here, $\log_{10}U=1890$) are low-noise instanton realizations. This is in agreement with the instanton analysis which predicts noiseless instanton realizations in the limit $\ln U\to +\infty$.

To summarize, the numerical results of Ref.~\onlinecite{Mounaix2024} confirm the predictions made by the instanton analysis, both regarding the tail of $p(U)$ -- which determines the value of the critical couplig $g_c(L)$ -- and the corresponding realizations of $S(x,z)$. Since these realizations take the form of elongated, filament-like instantons rather than localized hot spot fields, the results of Secs.~\ref{inst} and~\ref{test} clearly show that instanton analysis rules out the Rose\textendash DuBois ansatz in the far upper tail of $p(U)$. The only exception occurs in the small length limit, $L\ll\Lambda_{\parallel}$, where instantons become indistinguishable from hot spots, up to second order in $L/\Lambda_{\parallel}$.
\begin{figure}[!b]
\includegraphics[width=0.9\linewidth]{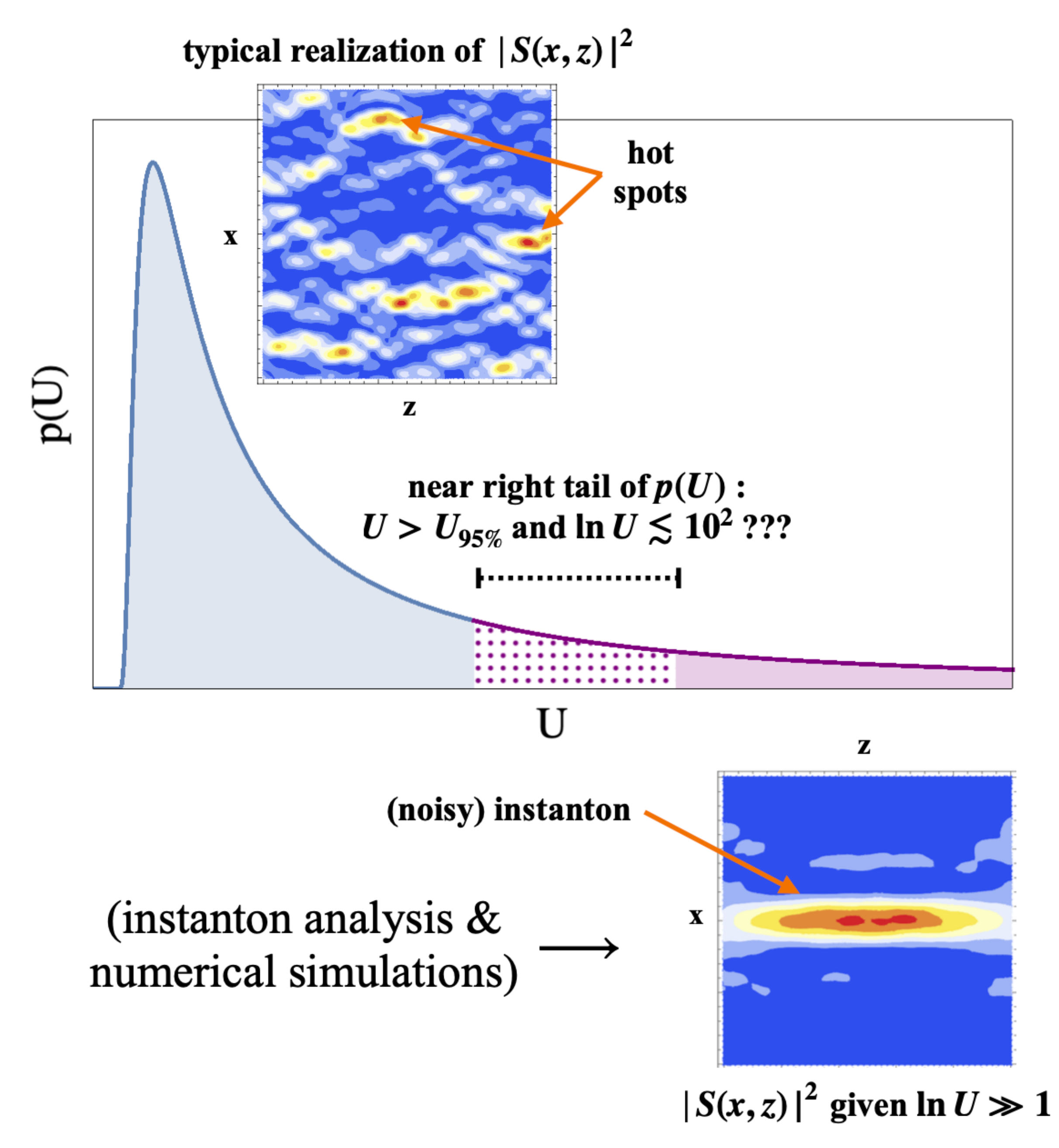}
\caption{\label{figure11}First updated version of Fig.~\ref{figure3}, incorporating the instanton analysis from Ref.~\onlinecite{Mounaix2023} and the numerical results from Ref.~\onlinecite{Mounaix2024}.}
\end{figure}

Figure~\ref{figure11} takes stock of what has been learned so far about the realizations of $S(x,z)$ as a function of $U$. Comparing with Fig.~\ref{figure3}, it can be seen that in the far upper tail of $p(U)$, the high-intensity hot spot assumed by the Rose\textendash DuBois ansatz must be replaced by a high-intensity instanton. That said, the story does not end there. Between the hot spot and instanton domains -- corresponding respectively to the bulk and the far upper tail of $p(U)$ -- there remains an entire intermediate zone, referred to as the near upper (or right) tail of $p(U)$, where the nature of the realizations of $S(x,z)$ is still unknown.

The study of this intermediate (or transition) region, say between the 95th percentile, $U_{95\%}$ and $\ln U\sim 10^2$, is particularly important, as it corresponds to an amplification range that may be directly relevant to laser-plasma interaction. In realistic physical situations, the far upper tail of $p(U)$ is practically unreachable -- but the near upper tail {\it is} accessible. It is therefore crucial to understand the behavior of the realizations of $S(x,z)$ in this regime. This is the focus of the next section.
%
%
\section{Transition regime and complete scenario}\label{trans}
\subsection{Numerical results in the near upper tail of $\bm{p(U)}$}\label{numericsneartail}
In the transition regime, no systematic analytical method, like instanton analysis, is available. It follows that numerical simulations are the only way to gather information about the realizations of $S(x,z)$ in the near upper tail of $p(U)$. Unlike the far upper tail, the near upper tail can be accessed through standard direct sampling with reasonably large sample sizes. By {\it reasonably large}, we mean that the sample size must be large enough to probe amplification values deep enough within the near upper tail of $p(U)$, without requiring prohibitively long computation times. Here, we present the main results of the numerical study reported in Ref.~\onlinecite{Mounaix2023} for a sample of $10^5$ independent realizations of $S(x,z)$. The parameters and the Gaussian field $S$ are the same as in Sec.~\ref{numericsfartail}. In this setting, amplification values as large as $\ln U\simeq 65$ in the near upper tail can be probed, providing valuable insight into the transition region between the bulk and the far upper tail of $p(U)$, as we will now see.

To get better statistics of large amplifications, it is useful to consider the maximum amplification $U_{\rm max}=\max_x\vert\psi(x,L)\vert^2$ rather than $U=\vert\psi(0,L)\vert^2$. In addition to its practical benefit, considering $U_{\rm max}$ is also physically motivated. This motivation arises from the intermittent behavior briefly discussed in Sec.~\ref{boflit}. Write $\mathcal{P}=\mathcal{P}_{th}\, \overline{U}$ the power of the backscattered light, with thermal level $\mathcal{P}_{th}$, and where $\overline{U}=\ell^{-1}\int_{-\ell/2}^{\ell/2}\vert\psi(x,L)\vert^2\, dx$ is the space average of the amplification. $\overline{U}$ plays the same role as the sample average in Sec.~\ref{boflit}. As explained there, in the supercritical regime, $g>g_c(L)$, and for $\ell/\Lambda_{\perp}$ large enough, $\mathcal{P}$ is determined by the contribution of $\vert\psi(x,L)\vert^2$ near its maximum. More precisely, for every realization of $S(x,z)$ one has $\mathcal{P}\sim\Delta x\, U_{\rm max}$ when $\ell/\Lambda_{\perp}\gg 1$, where $\Delta x$ is a slowly varying effective width depending on the logarithm of $U_{\rm max}$. It can be shown\cite{EKM1997,RT2001} that the tail of $p(U_{\rm max})$ has a dominant algebraic part which is the same as that of $p(U)$ in Eq.~(\ref{logfartailofpofu}). This implies in particular that both $\langle U_{\rm max}\rangle$ and $\langle\mathcal{P}\rangle$ have the same critical coupling as $\langle U\rangle$, as confirmed by the evident equality $\langle\mathcal{P}\rangle =\mathcal{P}_{th}\, \langle U\rangle$.

Since, in principle, $U_{\rm max}$ is not the amplification at $x=0$, the instanton to be compared with $S(x,z)$ when $U_{\rm max}$ is large is no longer constrained to arrive at $x_{\rm inst}(L)=0$. Instead, its location along $x$ is now a free parameter that can be varied. For each realization, $S(x,z)$ must be compared with the closest instanton, in the sense of the same $L^2$-distance $\mathfrak{D}$ as in Fig.~\ref{figure9}, with minimum distance $\mathfrak{D}_{\rm min}$. In Ref.~\onlinecite{Mounaix2023}, $U_{\rm max}$ and $\mathfrak{D}_{\rm min}$ have been determined numerically for each of the $10^5$ realizations of the sample. Then the behavior of the statistics of $\mathfrak{D}_{\rm min}$ as a function of $U_{\rm max}$ has been studied, with the results that we will now present.
\begin{figure}[!b]
\includegraphics[width=1.0\linewidth]{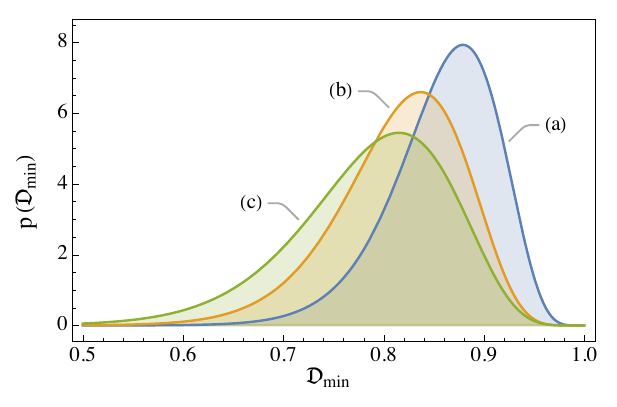}
\caption{\label{figure12}Probability distribution of $\mathfrak{D}_{\rm min}$ estimated from the $10^5$ randomly drawn realizations of $S(x,z)$ with (a) no restriction on $U_{\rm max}$, (b) $U_{\rm max}>5\times 10^{12}$ (90th percentile), and (c) $U_{\rm max}>10^{18}$ (99th percentile). The most probable value is $U_{\rm max}\simeq  10^6$. (Fig. 2 of Ref.~\onlinecite{Mounaix2023}. Copyright 2023 IOP Publishing Ltd)}
\end{figure}

Figure~\ref{figure12} shows the probability distribution of $\mathfrak{D}_{\rm min}$ estimated from the $10^5$ randomly drawn realizations of $S(x,z)$ with (a) no restriction on $U_{\rm max}$, (b) $U_{\rm max}>5\times 10^{12}$ (90th percentile), and (c) $U_{\rm max}>10^{18}$ (99th percentile). For comparison, the most probable amplification value estimated from the same sample is $U_{\rm max}\simeq  10^6$ (see Fig.~3 of Ref.~\onlinecite{Mounaix2023}). As can be seen, the distribution of $\mathfrak{D}_{\rm min}$ shifts to the left from (a) to (c), indicating a tendency for $\mathfrak{D}_{\rm min}$ to decrease with increasing $U_{\rm max}$. As a side note, the difference between $p(\mathfrak{D})$ in the inset of Fig.~\ref{figure9} and $p(\mathfrak{D}_{\rm min})$ in Fig.~\ref{figure12}, curve (a), arises from the fact that, for almost all realizations, $\mathfrak{D}_{\rm min}<\mathfrak{D}\le 1$. This causes a drop in the number of realizations with $\mathfrak{D}_{\rm min}$ close to the maximum value $1$, compared to the corresponding number for $\mathfrak{D}$.
\begin{figure}[!b]
\includegraphics[width=1.0\linewidth]{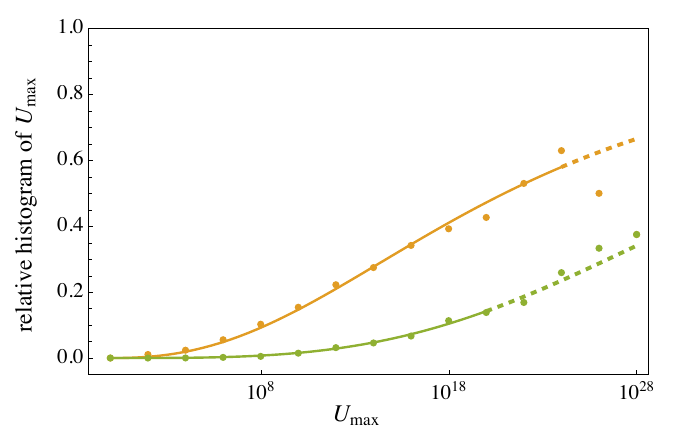}
\caption{\label{figure13}Percentage of realizations with $\mathfrak{D}_{\rm min}<0.8$ (orange, upper curve) and $\mathfrak{D}_{\rm min}<0.7$ (green, lower curve) for $\log_{10}U_{\rm max}$ in $\lbrack n,n+2)$, with $n$ varying from $0$ to $28$. $\mathfrak{D}_{\rm min}=0.8$ and $\mathfrak{D}_{\rm min}=0.7$ correspond to the 10th and 1st percentiles. (Fig. 4 of Ref.~\onlinecite{Mounaix2023}. Copyright 2023 IOP Publishing Ltd)}
\end{figure}

Figure~\ref{figure13} reveals the same phenomenon as Fig.~\ref{figure12}, but in a different way. It shows the fraction of realizations for which $\mathfrak{D}_{\rm min}$ is less than a given value, as a function of $U_{\rm max}$. These results have been obtained by counting the realizations of $U_{\rm max}$ in bins of two decades in width, then computing the fraction within each bin. The orange, upper curve corresponds to $\mathfrak{D}_{\rm min}<0.8$ (10th percentile) and the green, lower curve to $\mathfrak{D}_{\rm min}<0.7$ (1st percentile). It can be seen that both curves increase with increasing $U_{\rm max}$, reflecting the same bias of $S$ toward the instanton as observed in Fig.~\ref{figure12}. Consider the lower, green curve, for example. For $U_{\rm max}$ between $10^7$ and $10^9$, the weight of the realizations with $\mathfrak{D}_{\rm min}<0.7$ is negligible, whereas for $U_{\rm max}$ between $10^{27}$ and $10^{29}$, it rises to about $30\%$. This is not yet dominant, but definitely no longer negligible. If $U_{\rm max}$ were increased further -- e.g., by increasing the sample size -- the curve would continue to rise, eventually approaching $100\%$ upon entering the asymptotic regime, where the predictions of Secs.~\ref{inst} and \ref{test} are realized.
\begin{figure*}[!t]
\includegraphics[width=0.75\linewidth]{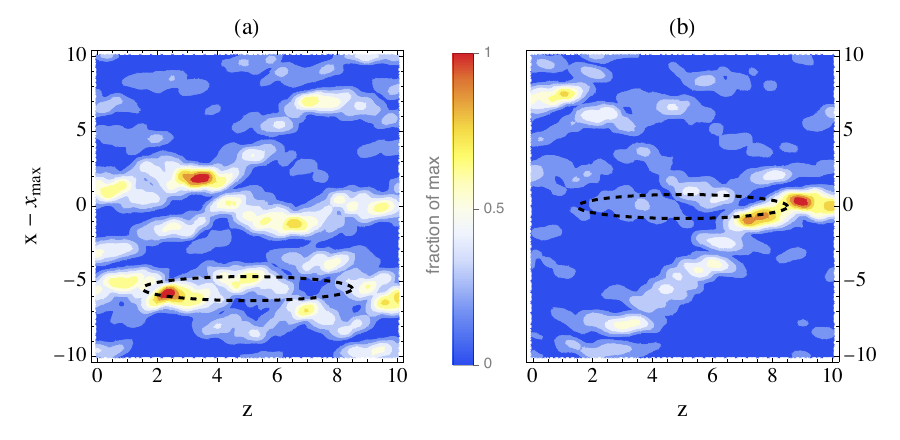}
\caption{\label{figure14}Contour plots of $\vert S(x,z)\vert^2$ for two typical realizations with $\mathfrak{D}_{\rm min}>0.7$ (1st percentile). $x_{\rm max}$ is the position of the maximum of $\vert\psi(x,L)\vert^2$. The dashed contours indicate the closest theoretical instanton, at distance $\mathfrak{D}_{\rm min}$ from the corresponding $S$ profile. (a) $\log_{10}U_{\rm max}=29.84$, $\mathfrak{D}_{\rm min}=0.8$, and (b) $\log_{10}U_{\rm max}=27.69$, $\mathfrak{D}_{\rm min}=0.82$. (Fig. 7 of Ref.~\onlinecite{Mounaix2023}. Copyright 2023 IOP Publishing Ltd)}
\end{figure*}
\begin{figure*}[!t]
\includegraphics[width=0.75\linewidth]{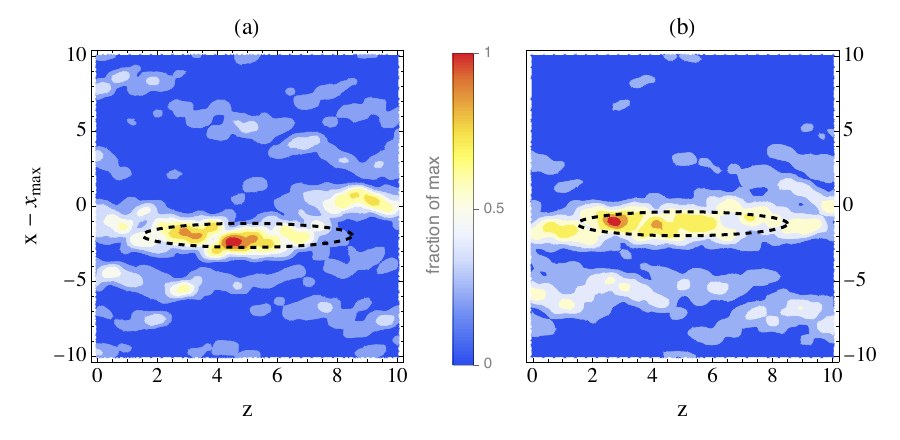}
\caption{\label{figure15}Plots similar to the ones shown in Fig.~\ref{figure14} for two typical realizations with $\mathfrak{D}_{\rm min}\le 0.7$ (1st percentile). (a) $\log_{10}U_{\rm max}=29.63$, $\mathfrak{D}_{\rm min}=0.68$ and (b) $\log_{10}U_{\rm max}=27$, $\mathfrak{D}_{\rm min}=0.537$. (Fig. 6 of Ref.~\onlinecite{Mounaix2023}. Copyright 2023 IOP Publishing Ltd)}
\end{figure*}

\begin{figure*}[!t]
\includegraphics[width=0.66\linewidth]{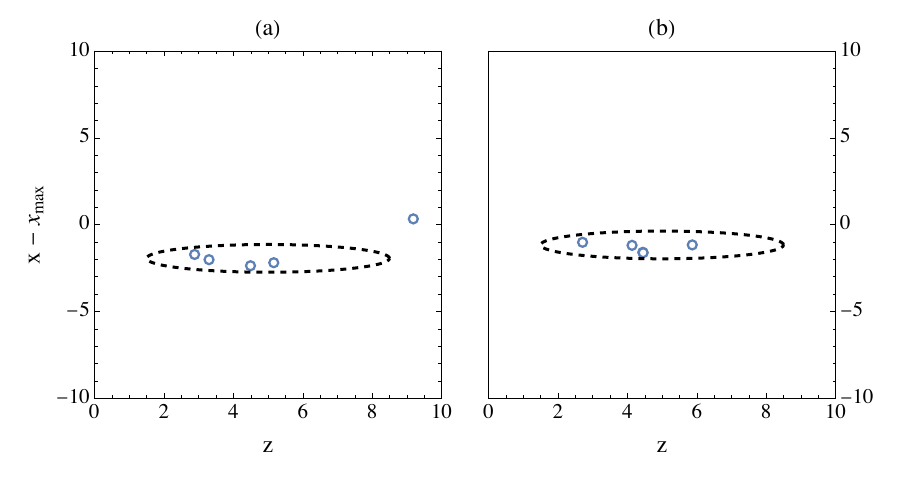}
\caption{\label{figure16}Positions of the local maxima of $\vert S(x,z)\vert^2$ higher than $75\%$ of the global maximum (blue circles) for the same realizations as in Fig.~\ref{figure15}. High maxima cluster in the instanton region indicated by the dashed contour. (See Fig.~\ref{figure15} for details.) (Fig. 10 of Ref.~\onlinecite{Mounaix2023}. Copyright 2023 IOP Publishing Ltd)}
\end{figure*}

The results of Figs.~\ref{figure12} and \ref{figure13} clearly indicate a tendency for the realizations of $S(x,z)$ to get closer to the instanton as the amplification increases. It now remains to understand how this tendency manifests in the realizations of $S(x,z)$ themselves. Figures~\ref{figure14} and \ref{figure15} display realizations of $\vert S(x,z)\vert^2$ for  $U_{\rm max}$ between $10^{27}$ and $10^{29}$. Intense localized hot spots similar to the theoretical one in Fig.~\ref{figure7}(b) are visible in both figures. Two typical realizations from the $70\%$ with $\mathfrak{D}_{\rm min}>0.7$ (1st percentile) are shown in Fig.~\ref{figure14}. Hot spots appear uniformly scattered throughout the box, and the aspect of $\vert S(x,z)\vert^2$ is similar both inside and outside the nearest instanton region, indicated by a dashed contour. The fraction of laser energy in the instanton component is insufficient for the instanton to be distinctly visible, which is consistent with $\mathfrak{D}_{\rm min}$ being close to $1$. In summary, these realizations resemble the hot spot fields assumed by the Rose\textendash DuBois ansatz. Figure~\ref{figure15}, on the other hand, shows two typical realizations from the $30\%$ with $\mathfrak{D}_{\rm min}\le 0.7$ (1st percentile). Hot spots appear clustered inside the dashed contour, within the nearest instanton region, in striking contrast to the dispersed hot spots observed in Fig.~\ref{figure14}. Note also that the level of $\vert S(x,z)\vert^2$ is significantly higher than average throughout the instanton region ($\sim 6$, while $\langle\vert S(x,z)\vert^2\rangle=1$), which is difficult to explain by generic Gaussian fluctuations (i.e. independent, small-scale hot spots). Hot spot clustering is clearly visible in Fig.~\ref{figure16}, which focuses on hot spot positions for the same realizations as in Fig.~\ref{figure15}. We will call the instanton together with its surrounding cluster of hot spots an {\it instanton\textendash hot spot complex}. From the above discussion of the numerical results of Ref.~\onlinecite{Mounaix2023}, it follows that the realizations of $S(x,z)$ underlying the right part of the lower, green curve in Fig.~\ref{figure13} are characterized by the presence of instanton\textendash hot spot complexes, which are absent from the hot spot field description of Rose and DuBois. The emergence of such complexes in a non-negligible fraction of realizations is a remarkable feature of $S(x,z)$ in the near upper tail of $p(U)$.

Taken together, the results from Sec.~\ref{theinstsol} to Sec.~\ref{numericsneartail} yield a comprehensive picture of how the realizations of $S(x,z)$ depend on the amplification. The next part is dedicated to describing this dependence in more detail.
\subsection{Complete scenario}\label{completescenar}
\begin{figure}[!b]
\includegraphics[width=0.9\linewidth]{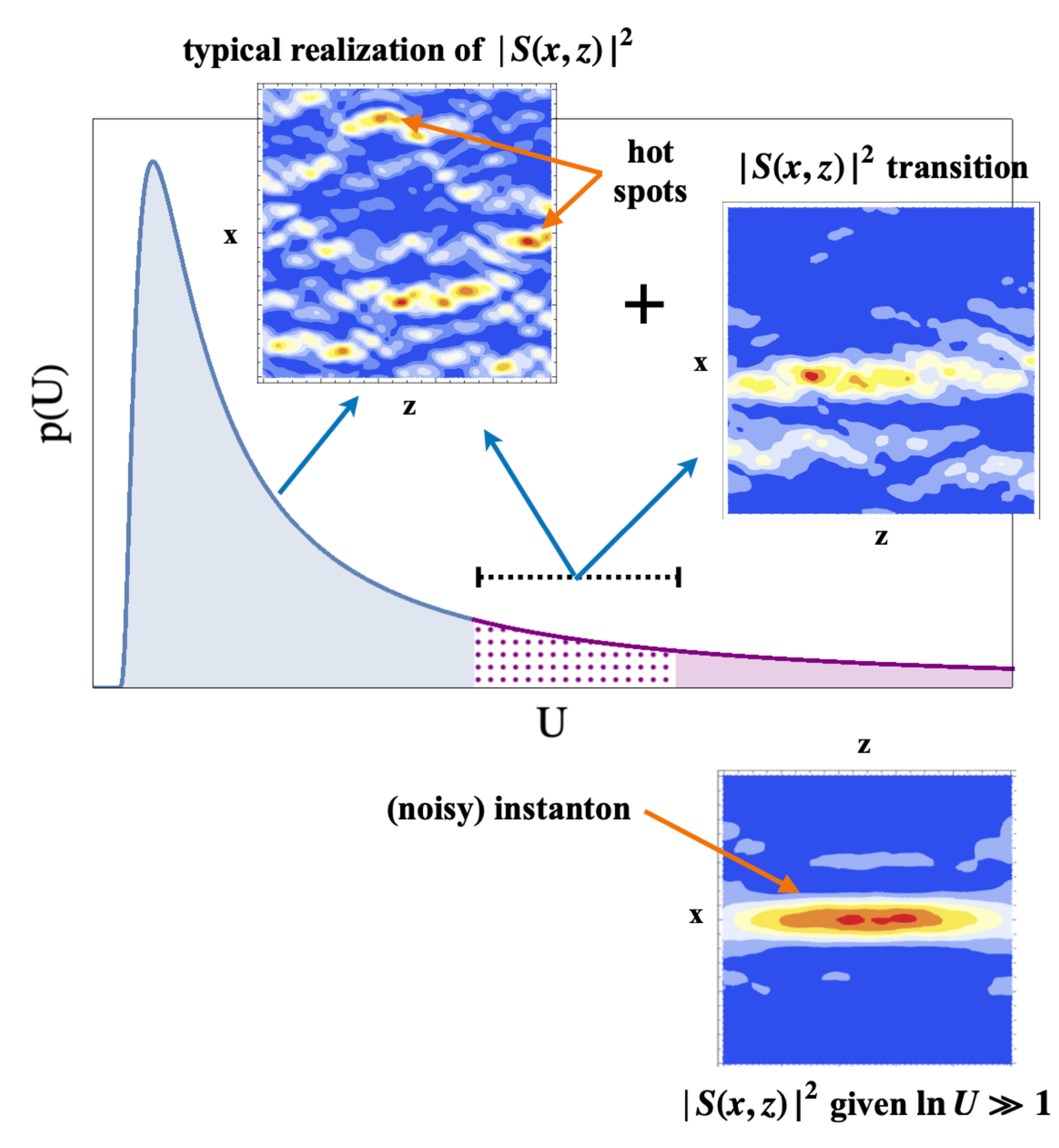}
\caption{\label{figure17}(Final updated version of Fig.~\ref{figure3}) Representative realizations of $\vert S(x,z)\vert^2$ as a function of the region of $p(U)$: bulk (left); near upper tail (center); far upper tail (right).}
\end{figure}

Figure~\ref{figure17} offers a clear, easy-to-grasp summary of the emerging scenario and its differences from the Rose\textendash DuBois ansatz represented in Fig.~\ref{figure3}.

\smallskip
{\it (i) $U$ in the bulk of $p(U)$}. For the most probable values of $U$ -- say, $U\lesssim U_{95\%}$ -- the realizations of $S(x,z)$ are hot spot fields. In this regime, Figs.~\ref{figure3} and \ref{figure17} are identical.

\smallskip
{\it (ii) $U$ in the near upper tail of $p(U)$}. For large values of $U$ within the near upper tail of $p(U)$ -- specifically, $U\gtrsim U_{95\%}$ and $\ln U\lesssim 10^2$ -- we observe a coexistence of hot spot fields and instanton\textendash hot spot complexes, with relative population sizes depending on the value of $U$, and all intermediate configurations possible in terms of how the laser energy is distributed between the hot spots and the instanton. (a) On the left side of the near upper tail, most realizations are hot spot fields, with most of the laser energy concentrated in the hot spots and little in the instanton. (b) As $U$ increases, hot spot fields see their majority decline, while the population of instanton\textendash hot spot complexes grows. This change is accompanied by an increase of the laser energy in the instanton, at the expense of the hot spots. (c) On the right side of the near upper tail, most realizations are instanton\textendash hot spot complexes, with the instanton carrying more laser energy than the hot spots.

\smallskip
{\it (iii) $U$ in the far upper tail of $p(U)$}. Finally, for even larger values of $U$, we enter the asymptotic regime that defines the far upper tail of $p(U)$ -- that is, $\ln U\gg 1$. Here, almost all realizations are instantons, with virtually no laser energy remaining in the hot spots, which appear as background noise (for finite $\ln U$). This situation is illustrated in the bottom-right corner of Fig.~\ref{figure17}, where the noise superimposed on the instanton is the remnant of the hot spots in the asymptotic regime.

\smallskip
What Figure~\ref{figure17} reveals is a complete reversal of roles between hot spots and instantons when moving from the bulk to the far upper tail of $p(U)$. This is in striking contrast to Fig.~\ref{figure3}, where only hot spots are present -- by assumption of the Rose\textendash DuBois ansatz. It follows in particular that the divergence of $\langle U\rangle$ for $g>g_c(L)$ is caused by high-intensity instantons, not by high-intensity hot spots. (Except in the small length limit, $L\ll\Lambda_{\parallel}$, where instantons and hot spots coincide up to second order in $L/\Lambda_{\parallel}$.)

Finally, it is important to note that, according to the numerical results of Refs.~\onlinecite{Mounaix2023,Mounaix2024}, instanton\textendash hot spot complexes account for $5$ to $10\%$ of the realizations at the lower limit of the transition regime -- the near upper tail of $p(U)$ -- and up to nearly $100\%$ of the realizations in the asymptotic regime \mbox{-- the far upper tail of $p(U)$ --}, with a monotonic growth as $U$ increases. They are therefore never entirely negligible -- and should not be overlooked -- throughout the full extent of the upper tail of $p(U)$, unlike in the bulk, where their contribution is entirely negligible.
%
%
\section{Potential consequences for laser-plasma interaction: preliminary considerations}\label{conse}
In this last section before the conclusion, we offer preliminary considerations regarding potential implications of the preceding results for laser-plasma interaction. The aim is not to address the questions raised by these implications, each of which constituting a research topic in its own right, but simply to identify them in the hope that they will be explored in future work. Accordingly, and to maintain a pedagogical tone, we will present these implications in qualitative terms, without delving into quantitative specifics.

\subsection{Expressing $\bm{g}$ and $\bm{m}$ as functions of the physical parameters}\label{gandmphysics}
For practical applications, it may be helpful to make explicit the connection between the constants $g$ and $m$ appearing in Eq.~(\ref{withDeq}) and the parameters that govern these quantities, such as laser intensity, temperatures(s), damping, density etc. One has
\begin{equation}\label{gandmexpressions}
g=\frac{\gamma_0^2}{V_1\nu_2}\ {\rm and}\ m=k_1,
\end{equation}
where $V_1$ and $k_1$ respectively denote the group velocity and wave number of the backscattered wave, and $\nu_2$ is the linear damping of the electrostatic wave (acoustic wave for Brillouin scattering, plasma wave for Raman scattering). The coupling parameter $\gamma_0$ is the growth rate of the instability at average laser intensity. In the case of backward stimulated Brillouin scattering, $\gamma_0$ is given by
\begin{equation}\label{gamma0sbs}
\gamma_0({\rm ps}^{-1})=\frac{4.14(Z/A)^{1/4}(n/n_c)^{1/2}I_{14}^{1/2}}{T_e^{1/4}(1+k_2^2\lambda_D^2)^{3/4}\lbrack 1+\tau(1+k_2^2\lambda_D^2)\rbrack^{1/4}},
\end{equation}
with $\tau=3T_i/ZT_e$ and $k_2^2\lambda_D^2=7.83\times 10^{-3}T_e(n_c/n-1)$, where $k_2$ is the wave number of the acoustic wave. The quantities $T_e$ and $T_i$ denote the electron and ion temperatures in keV, $n_c$ is the critical density, $I_{14}$ is the average laser intensity in $10^{14}\ {\rm W}\ {\rm cm}^{-2}$, and $Z$ (resp. $A$) is the ion charge (resp. mass) number. The backscattered wave is characterized by
\begin{equation}\label{k1sbs}
k_1\simeq\frac{2\pi}{\lambda_0}\left(1-\frac{n}{n_c}\right)^{1/2},
\end{equation}
and
\begin{equation}\label{V1sbs}
V_1\simeq c\left(1-\frac{n}{n_c}\right)^{1/2},
\end{equation}
where $\lambda_0$ is the laser wavelength in vacuum and $c$ is the speed of light. The acoustic wave is characterized by $k_2\simeq 2k_1$ and $\nu_2$ computed from, e.g., the convenient analytical fits (taking into account both Landau and collisional effects) proposed in Ref.~\onlinecite{Casanova1989}. In the case of backward stimulated Raman scattering, $\gamma_0$ is given by
\begin{equation}\label{gamma0srs}
\gamma_0({\rm ps}^{-1})=\frac{8.05(n/n_c)^{1/4}I_{14}^{1/2}}{(1-n/n_c)^{1/4}}\sqrt{\frac{\omega_0}{\omega_1}}\frac{ck_2}{2\omega_0},
\end{equation}
where $\omega_0=2\pi c/\lambda_0$ and
\begin{equation}\label{k2srs}
k_2=\frac{\omega_0}{c}\left(1-\frac{n}{n_c}\right)^{1/2}+k_1,
\end{equation}
is the wave number of the plasma wave. The backscattered wave is characterized by
\begin{eqnarray}\label{k1srs}
\frac{ck_1}{\omega_0}&=&\left\lbrack 1-2\sqrt{\frac{n}{n_c}}-\frac{3T_e}{511}\left(\sqrt{\frac{n_c}{n}}-1\right)\right. \nonumber \\
&&\times\left.\left(\sqrt{1-2\left(\frac{n}{n_c}\right)^{1/2}}+\sqrt{1-\frac{n}{n_c}}\right)^2\right\rbrack^{1/2},
\end{eqnarray}
and
\begin{equation}\label{omega1srs}
\frac{\omega_1}{\omega_0}=1-\left(\frac{n}{n_c}\right)^{1/2}\sqrt{\frac{1+3k_2^2\lambda_D^2}{1+0.1(k_2\lambda_D)^{3/2}}},
\end{equation}
with $k_2\lambda_D=4.5\times 10^{-2}(ck_2/\omega_0)(n_cT_e/n)^{1/2}$. Convenient analytical fits for the computation of the linear damping of the plasma wave $\nu_2$ can also be found in Ref.~\onlinecite{Casanova1989}.
%
%
\subsection{Preliminary discussion of potential consequences}\label{outlinecons}
To begin with, we discuss the appropriateness of the periodic boundary conditions in the transverse direction, as used in Refs.~\onlinecite{Mounaix2023,Mounaix2024} (as well as in Ref.~\onlinecite{MCL2006}). From a physical standpoint, open (or absorbing) boundary conditions may be more realistic for simulating laser-plasma interactions, potentially affecting the formation and confinement of instanton structures. In the limit where the width of the interaction region is much larger than the transverse extent of the instanton, defined as the distance spanned by the instanton in the transverse direction, boundary effects are negligible and the choice of boundary conditions is irrelevant. For widths larger than the instanton extent, but not much larger, all instanton realizations that cross the boundaries are suppressed (or strongly affected) if open (or absorbing) boundary conditions are chosen. In this case, the conclusions of Refs.~\onlinecite{Mounaix2023,Mounaix2024} still apply, but with instantons confined to the central part of the interaction region, where they do not cross the boundaries. The case of an interaction region with open (or absorbing) boundary conditions and a width smaller than the instanton extent requires specific treatment, as the instanton solutions obtained in Refs.~\onlinecite{Mounaix2023,Mounaix2024} are all suppressed (or strongly affected). Consequently, the conclusions of Refs.~\onlinecite{Mounaix2023,Mounaix2024} still apply for open (or absorbing) boundary conditions, as long as the width of the interaction region exceeds at least a few instanton extents. For the instantons considered in Sections~\ref{inst} and \ref{test}, this means $\ell$ greater than at least a few $\Lambda_{\perp}$. (Measured and simulated focal spots at large laser facilities\cite{Lindl2004,Haynam2007} typically exhibit $\ell/\Lambda_\perp$ ratios between $10^2$ and $10^3$.)

In laser-plasma interaction, scattering instabilities are commonly characterized by the reflectivity, $R$, defined as the (physical) intensity of the backscattered light normalized to the average laser intensity. Therefore, it is useful to specify the relationship between $R$ and the amplification $U_{\rm max}$. Physically, the scattering process is seeded by thermal fluctuations of light and acoustic waves in the plasma, with the contribution of the latter usually exceeding that of the former (the reasoning is the same with non-thermal fluctuations). In the stationary convective regime, the plasma thermal noise can be modeled by the boundary condition $\psi(x,0)=\psi_{th}$, where the constant $\psi_{th}>0$ represents an effective thermal level that depends on both the acoustic and light plasma noises. Note that normalizing $\psi$ to $\psi_{th}$ yields the boundary condition $\psi(x,0)=1$, as used in Refs.~\onlinecite{Mounaix2023,Mounaix2024} (see above Eq.~(\ref{generalactionint})). In the linear, supercritical regime ($g>g_c(L)$), $R$ and $U_{\rm max}$ are related by $R\simeq R_{th}\, U_{\rm max}$, where $R_{th}\propto\psi_{th}^2$ is the corresponding thermal level of $R$ ($R_{th}$ is the Thomson scattering reflectivity off the thermal fluctuations of the plasma). For example, taking a typical value $R_{th}\sim 10^{-9}\text{--}10^{-10}$, a range of $U_{\rm max}$ between $10^{17}$ and $10^{26}$ corresponds to $R$ ranging from $10^{7}\text{--}10^{8}$ up to $10^{16}\text{--}10^{17}$, which gives an idea of the (linear) reflectivities involved.

Such large values of $R$ are clearly unrealistic, which is not surprising for a linear amplifier like Eq.~(\ref{withDeq}) which lacks any saturation mechanism. The same applies to the numerical results of Rose and DuBois reproduced in Fig.~\ref{figure2}, where the divergence of $\langle U\rangle$ is inferred from unphysically large values of $U$. The relevance of a linear approach to a realistic description of the problem lies in the fact that Eq.~(\ref{withDeq}) serves only as a preliminary step -- not yet intended to provide a fully realistic description. Rather, Eq.~(\ref{withDeq}) should be understood as a linear background that enables the unambiguous definition of the critical coupling and the identification of the structures of $S$ that contribute to amplification. Nonlinear effects are then incorporated into these structures to provide a more realistic description in the supercritical regime, where Eq.~(\ref{withDeq}) locally breaks down within the dominant structures\cite{RD1994}, yielding physically meaningful values of $R$.

The conditions under which Eq.~(\ref{withDeq}) can be used as the appropriate linear background have been discussed in Ref.~\onlinecite{RD1994}. Without going into detail, these conditions require that (i) linear backscattering within the dominant structures remains below the threshold for the onset of absolute instability\cite{PLP1973,Briggs1964,Bers1983,Kroll1965}, and (ii) the stationary convective regime is reached within the duration of the laser pulse. The latter can be assessed using the dynamical theory of backscattering instabilities\cite{MPRC1993,MP1994,HWB1994,DM1999a,DM1999b}. (Readers interested in quantitative details are referred to the relevant references.) Under these conditions, the reasoning developed by Rose and DuBois in Ref.~\onlinecite{RD1994} applies, but with the important difference that the dominant structures of $S$ do not amount  to the hot spots represented in Fig.~\ref{figure3}, as assumed by Rose et DuBois, but instead correspond to the structures shown in Fig.~\ref{figure17}. The implication of the work in Refs.~\onlinecite{Mounaix2023,Mounaix2024} for laser-plasma interaction is that Fig.~\ref{figure3} must be replaced by Fig.~\ref{figure17} in the reasoning of Rose and DuBois.

Given the inherent limitations of the statistics of $U$ in all practical situations, the far upper tail of $p(U)$ -- where pure instantons reside -- appears to be physically inaccessible, as noted at the end of Section~\ref{test}. In this respect, the extreme nature of the instanton solution may render it more of a theoretical construct than a physically meaningful quantity, even though it entirely determines the value of the critical coupling $g_c(L)$. The physically meaningful part of Fig.~\ref{figure17} corresponds to the near upper tail of $p(U)$, which is accessible in physical situations. This is where the results of Ref.~\onlinecite{Mounaix2023} may modify the Rose and DuBois analysis, as we now explain.

As suggested by the final remark of Section~\ref{trans}, the hot spot field description of Rose and DuBois should be supplemented to include the possibility of realizations of $S$ involving instanton\textendash hot spot complexes, in addition to the expected hot spot field realizations. Among the realizations that dominate the power of the backscattered light in the supercritical regime ($g>g_c(L)$), the fraction involving instanton\textendash hot spot complexes is {\it not} negligible, contrary to the implicit assumption made by Rose and DuBois in their 1994 paper\cite{RD1994}. It is important to emphasize that, within the experimentally accessible range of large $U$, realizations of $S$ involving instanton\textendash hot spot complexes do not constitute the majority. Accordingly, replacing Fig.~\ref{figure3} with Fig.~\ref{figure17} does not invalidate the hot spot field description of laser-plasma interaction; rather, it complements it by accounting for the presence of additional structures beyond hot spots.

Following these observations, the concept of instanton\textendash hot spot complexes should be incorporated into the theoretical toolbox of laser-plasma interaction, particularly in cases where large-scale filamentary structures in the laser field are suspected. For specialists familiar with concrete realizations of laser light fields generated by algorithms mimicking RPP (or similar optical smoothing methods) in the focal domain, the observation of elongated structures at non-negligible intensity, sometimes directly or obliquely connecting hot spots, is certainly not surprising. When exploring different realizations of RPP masks, one might indeed observe that the appearance of such interconnecting structures between hot spots is not so rare. The fact that the intensity does not significantly drop between neighboring hot spots can have important consequences for the amplification of scattered light. High amplification does not necessarily result from localized, high-intensity hot spots; rather, it may arise from structures that are less intense but significantly longer, such as instanton\textendash hot spot complexes -- thereby motivating further study of such structures in the context of laser-plasma interaction.

Up to now, the emergence of interconnecting structures between hot spots can be explained as the outcome of the plasma-induced evolution of an initially unstructured hot spot field -- ``initially'' here referring to the linear background given by Eq.~(\ref{withDeq}). In this view -- which corresponds to the Rose\textendash DuBois ansatz of hot spot fields -- small-scale, initially uncorrelated hot spots mutually interact through plasma effects, primarily ion-acoustic waves, eventually forming a structure of interconnected, correlated hot spots. Note that plasma-induced correlations between hot spots was acknowledged in Ref.~\onlinecite{RD1994} as a possibility and have remained an ongoing concern in subsequent applications of the hot spot model (see, e.g., Refs.~\onlinecite{RG1998,RLT2024}).

The results in Ref.~\onlinecite{Mounaix2023} offer an unexpected alternative explanation, based on the evolution of pre-existing large-scale structures -- the instanton\textendash hot spot complexes -- already present in the linear background. In this scenario, hot spot correlations result from the conditioning of the laser field to a large value of the amplification, through the emerging instanton structure. To distinguish them from the plasma-induced correlations mentioned above, we refer to them as {\it instanton-induced} correlations. In practice, instanton-induced correlations are likely accompanied by plasma-induced correlations, so that observed correlations should be a combination of both mechanisms, in proportions yet to be determined. Since the near upper tail of $p(U)$ is accessible with typical laser shot parameters, instanton-induced correlations could potentially be observed using current diagnostics (e.g., imaging of scattered light, beam imprints) through the identification of high-intensity hot spots clustered within or near the theoretical instanton region. By contrast, in the absence of instanton\textendash hot spot complexes, plasma-induced correlations alone have no reason to align hot spots along the theoretical instanton filament(s).

What is now needed is a statistical theory of $S(x,z)$ in the near upper tail of $p(U)$ (in this domain, only numerical results are available so far). As an improvement on the statistically independent hot spot model of Rose and DuBois, such a theory should account for both standard hot spot fields and instanton\textendash hot spot complexes, incorporating both plasma-induced and instanton-induced correlations between hot spots. This would make it possible to (i) assess how hot spot correlations impact the applicability and interpretation of the independent hot spot model, and (ii) delineate a concrete parameter space in which the relevance and/or impact of instanton\textendash hot spot complexes should be expected, both for near-backward Brillouin and Raman scattering. Readers accustomed to working with the independent hot spot model would thus benefit from substantial progress -- not only in relation to the more mathematical treatment presented in Ref.~\onlinecite{Mounaix2023}, but also with regard to phenomena observed in experiments or simulations that remain poorly understood to this day. Moreover, having such a theory would provide much stronger grounds for discussing the validity and limitations of the linear background approach in constructing a realistic nonlinear description of backscattering instabilities.

Developing and leveraging such a statistical theory remains an open task, far beyond the scope of this tutorial paper, the aim of which is limited to drawing the laser-plasma community's attention to the technical results of Refs.~\onlinecite{Mounaix2023,Mounaix2024}. This brings us to the final section, where we outline several directions for future investigation.
%
%
\section{Perspectives}\label{persp}
The results of Refs.~\onlinecite{Mounaix2023,Mounaix2024} reviewed in this tutorial represent only the very first step toward a comprehensive understanding of laser-plasma interaction with a smoothed laser beam in the supercritical regime, beyond the hot spot field description. There are many directions in which this investigation could be further pursued, not only in laser-plasma physics but also in mathematical and statistical physics. Focusing specifically on laser-plasma interaction, several avenues for future research can be identified.

As noted at the end of Sec.~\ref{outlinecons}, a important first follow-up is the development of a statistical theory for $S(x,z)$ in the near upper tail of $p(U)$, intended to quantify the coexistence of hot spots and instanton\textendash hot spot complexes, depending on the region of the parameter space. The goal is to construct a relatively simple yet realistic statistical model that enables quantitative predictions, while avoiding the otherwise inevitable need for intensive numerical simulations. With such a theory in hand, the next step is to assess the validity and limitations of the linear background approach, using Eq.~(\ref{withDeq}) without the Rose\textendash DuBois Ansatz, as a basis for constructing a realistic nonlinear description.

Another challenging line of research, complementing the previous one, is to go beyond the validity of Eq.~(\ref{withDeq}) as a linear background and give access to a broader range of physical situations. The first step in this direction is to introduce time dependence into the linear amplification of $\psi$, which allows for situations where the laser intensity may locally exceed the threshold for absolute instability, as well as interaction with a temporally smoothed laser beam. Performing the instanton analysis for a time-dependent stochastic amplifier is a complex task. In particular, it is highly unlikely that an analytical solution to the saddle-point equations can be found, as was the case in Ref.~\onlinecite{Mounaix2023} for the instanton analysis of Eq.~(\ref{withDeq}). Adapting an appropriate iterative forward-backward scheme, such as the one introduced in Ref.~\onlinecite{CS2001}, may provide a numerical solution to the saddle-point equations.

In the longer term, the final step toward a fully nonlinear description will involve incorporating nonlinearities from the outset, without relying on a linear background approach to first identify the structures on which nonlinear effects will develop. This nonlinear approach will necessitate the inclusion of an equation for $S$, which accounts for its depletion, filamentation, and self-focusing as it propagates through the plasma. In such a complex problem, the first challenge is to clearly define the supercritical regime, since the nonlinear saturation of amplification prevents any time-asymptotic divergence that would unambiguously define a critical coupling. Whether a supercritical regime does exist depends on how nonlinear effects suppress the upper tail of the amplification distribution. For a critical coupling to be definable, the tail must remain algebraic up to sufficiently large amplification values -- well beyond the bulk -- before being truncated by nonlinear effects. Given the complexity of the issue, only a comprehensive numerical study is likely to resolve the problem and fully clarify when and how instanton analysis remains relevant in the presence of nonlinearities.

Each of these research avenues constitutes a highly non-trivial topic in its own right that would merit further study. In conclusion, one could not do better than quote the final remark of Ref.~\onlinecite{Mounaix2023}: " it may be noted that the number of highly non-trivial questions raised by the seemingly simple linear problem (\ref{withDeq}) is quite remarkable. Following on from the work presented here, we hope that those questions will motivate interesting research [...]"
%
%
\section*{Data Availability Statement}
\vspace{-3mm}
The data that support the findings of this study are available within the article [and its supplementary material].
%
%

%
\end{document}